\documentclass{article}

\usepackage{amssymb}
\usepackage{amsmath}
\usepackage{amsmath,amsfonts,xypic}
\usepackage{amssymb}
\usepackage{amscd}
\usepackage{graphicx}

\begin{document}

\title{
{\bf  Geometrical foundations of plasticity yield criteria: A
unified theory }}

\author{
{\sf J. M. Luque}\thanks{Corresponding author: e-mail:
jmluque@us.es}$\,\,$${}^{a}$ {\sf R. Campoamor-Stursberg}
\thanks{e-mail:rutwig@pdi.ucm.es}$\,\,$${}^{b}$,
\\
{\small ${}^{a}${\it Dpto. Ingenier\'{\i}a Mec\'anica y de los
Materiales,}}\\
{\small {\it Escuela Superior de Ingenieros, Universidad de Sevilla,}} \\
{\small {\it Camino de los Descubrimientos s/n, 41902 Sevilla}}\\
{\small ${}^{b}${\it Instituto de Matem\'atica Interdisciplinar,}}\\
{\small {\it Universidad Complutense de Madrid, 3 Plaza de
Ciencias, 28040 Madrid, Spain.}}\\
}

\date{}

\maketitle

\begin{abstract}
A new model for elucidating the mathematical foundation of
plasticity yield criteria is proposed. The proposed ansatz uses
differential geometry and group theory concepts in addition to
elementary hypotheses based on well-established experimental
evidence. Its theoretical development involves the analysis of
tensor functions and provides a series expansion which allows the
functional stress-dependence of plasticity yield criteria to be
predicted. The theoretical framework for the model includes a
series of spatial coefficients that provide a more flexible theory
for in-depth examination of symmetry and anisotropy in compact
solid materials. It describes the classical yield criteria (like
those of Tresca, Von Mises, Hosford, Hill, etc) and accurately
describes the anomalous behaviour of metals such as aluminium,
which was elucidated by Hill (1979). Further, absolutely new
instances of stress-dependence are predicted; this makes it highly
useful for fitting experimental data with a view to studying the
phenomena behind plasticity.
\end{abstract}


\section{Introduction}

An accurate description of the structure, formation and behaviour
of a solid elastic-plastic material requires the knowledge, among
other facts, of the limiting stress it can withstand before it
becomes plastic [32, 33, 15, 18, 2, 22]. In fact, plasticity
concepts are widely used in a number of scientific and engineering
field applications (in materials science, physics of solids,
mechanical engineering, aeronautical engineering, geophysics,
biomechanics and chemistry, among others) [4, 8, 1].

Plasticity involves a series of irreversible, history-dependent
processes by effect of which a material develops fluency at a
micro-, meso- or macroscopic scale in its transition from an
elastic behaviour to a plastic behaviour [26]. Plastic processes
involve plastic dissipation (\textit{i.e.}, the irreversible
release of stress or energy with energy transfer in the material).
In fact, plasticity involves a variety of processes at different
spatial scales having an also different associated grain size. In
addition to its physical reality to plastic processes, grain size
defines characteristic spatial scales for macroscopic plasticity
in materials [19].

Unlike liquids and gases, solids are highly ordered systems,
contain a vast amount of internal information and exhibit a high
correlation among its constituent elements. As a result, plastic
microscopic processes in solids (movement of dislocations, defect,
etc) are usually relatively complex [19]. However, such processes
exhibit some macroscopic symmetry by virtue of the solids
structure and the physical laws they obey [4, 26, 5, 6]. This
facilitates the macroscopic examination of solids by using a
combination of differential geometry and group theory [3, 30].
Thus, a plastic process can be interpreted as a series of local
transformations that possess some symmetry and provide local
information useful with a view to establishing a global statistics
for a solid material.

The aim of this paper is to develop a unified approach of
plasticity yield criteria that uses elementary hypotheses based on
well-established experimental evidence. This study is based on the
analysis of the transformation properties of the Cauchy stress
tensor under orthogonal mappings, where arguments from the theory
of Lie groups are applied. Also, it discusses the physical
arguments for application of each criterion in relation to
specific properties of the material concerned. This can be useful
with a view to developing new criteria to address some special
mechanical properties of materials (like anisotropy,
hardening-softening, etc).

This paper is organized  as follows. Section 2 describes some
topics about plasticity in the framework of the internal variables
theory. Section 3 presents a short description of classical
plasticity criteria. In section 4 we show the postulates and
approximations of the unified theory. From section 5 onwards we
develop this unified theory. Finally, a set of conclusions is
shown.

\section{Plasticity function and plastic potential. Flow rule}

Within the framework of the Internal Variables Theory (or hidden
variables), an inelastic solid is one in which the strain at any
point of the solid is completely determined by the current stress
and temperature there plus a set of internal variables [26, 28].
The internal variables (scalars or tensors) have physical or
mathematical meaning and allow to complete the internal
description at any point of the solid (for example: the past
history of the stress and temperature at the point,
large-deformation plasticity, hardening and softening, structural
and induced anisotropy, etc) [26, 29]. Thus $\varepsilon $ is a
function of the material state $(\sigma ,T,\xi )$ at any
point\textbf{\underbar{}}

\noindent \textbf{\underbar{}}

\begin{equation} \label{GrindEQ__1_}
\varepsilon =\varepsilon (\sigma ,T,\xi )
\end{equation}

\noindent  where $\varepsilon $ denotes strain variables, $\sigma
$ stress variables, $T$ the temperature, and $\xi $ some internal
variables. Additionally, the rate of evolution of the internal
variables $\dot{\xi }$ is determined by the state

\noindent

\begin{equation} \label{GrindEQ__2_}
\dot{\xi }=\dot{\xi }(\sigma ,T,\xi )
\end{equation}

\noindent  known like equations of evolution or rate equation for
internal variables.

For inelastic solids  it is generally assumed that strain
variables can be decomposed additively into elastic strain
$\varepsilon ^{e} $ and inelastic strain $\varepsilon ^{i} $ [8]

\begin{equation} \label{GrindEQ__3_}
\varepsilon =\varepsilon ^{e} +\varepsilon ^{i}
\end{equation}

\noindent where the  inelastic strain occurring in
rate-independent plasticity is usually denoted by $\varepsilon
^{P} $rather than by $\varepsilon ^{i} $, and is called the
plastic strain.

In the context of internal  variables theory and rate-independent
plasticity, a plasticity yield criterion consists of a series of
mathematical conditions mutually relating stress, temperature and
internal variables, which define the material states $(\sigma
,T,\xi )$, where the point of  the solid concerned becomes
critically plastic. A critical state in this context is a state
where the elastic-plastic material starts to yield plasticity.
Therefore, these conditions constitute a boundary (plastic limit)
between the elastic and plastic state in the solid material (at
points). Mathematically, plasticity yield criteria are formulated
in the following general form:\textbf{\underbar{}}

\noindent

\begin{equation} \label{GrindEQ__4_}
f(\sigma ,T,\xi )=0
\end{equation}

\noindent where $f$ denotes  the plasticity function at a given
plastic state $(\sigma ,T,\xi )$. Therefore, the plasticity yield
criterion of a solid (at points) defines the stress multiaxial
states where it will yield critical plasticity; the set of such
states describes the criterion surface of the solid
\eqref{GrindEQ__4_}.

Plasticity can be defined  in basic terms by using some
approximations that usually hold in practice. Thus, the influence
of the temperature $T$ on solids at a constant, ambient level is
usually negligible provided they are scarcely sensitive to changes
in this variable and far from their melting point
(temperature-independent plasticity). The influence of strain is
usually negligible if we consider all viscoplastic processes to be
``infinitely'' slow compared to the material relaxation time
(rate-independent plasticity) [26]. Time-independent plasticity
needs usually to be considered if the target is not the
time-evolution of the solid. In the plastic limit, an index $P$ is
used to denote parameters and quantity values in such a limit. The
basic plastic unit in a solid is the macroscopic spatial point,
which can be equated to the physical concept of grain but need not
coincide with it. Also, because a solid usually exhibits high
correlation among its elements, characterizing each point in it
requires defining its correlation with its neighbourhood. As in
the theory of elasticity, this entails describing the stress state
at each point in terms of a second-order symmetric tensor
(2-tensor) $\sigma \equiv \sigma _{ij} $, called the
\textit{Cauchy tensor},\textit{ }the symmetry of which arises from
the stress equilibrium relation at the point in question [24].
Therefore, in these conditions, the plasticity yield criterion
\eqref{GrindEQ__4_} at each point in a solid can be defined as
follows:

\begin{equation} \label{GrindEQ__5_}
f(\sigma _{ij} ;\xi )=0
\end{equation}

\noindent which represents  the criterion surface in the stress
space.

The plastic potential, $\Phi $,  is a measure of smoothness
(differentiability) and convexity of $f$. Thus, if the plasticity
function $f$ is smooth (differentiable) and convex in stress
space, then it will coincide with a specific plastic potential
($f=\Phi $). The plasticity function $f=\Phi $ is convex if for a
given stress and strain rate condition $\sigma _{ij} $ and
$\dot{\varepsilon }_{ij}^{P} $, any stress $\sigma _{ij}^{*} $
inside or on the criterion surface obeys the following
relationship [26]:

\begin{equation} \label{GrindEQ__6_}
(\sigma _{ij}^{} -\sigma _{ij}^{*} )\dot{\varepsilon }_{ij}^{P} \ge 0
\end{equation}

\noindent where $\dot{\varepsilon }_{ij}^{P} $  denotes the
plastic strain rate. Specifically, if $f=\Phi $ is twice
differentiable, then $\Phi $ will be convex if, and only if, its
Hessian matrix, defined as

\noindent

\begin{equation} \label{GrindEQ__7_}
H_{ij} =\frac{\partial ^{2} \Phi }{\partial \sigma _{i} \partial \sigma _{j} }
\end{equation}

\noindent is positive semi-definite  (\textit{i.e. }if the
eigenvalues of $H_{ij} $ are only positive or zero). When the
eigenvalues are only positive non-zero (positive definite matrix),
$f=\Phi $ is strictly convex. We have used here ---as we have
throughout--- the Einstein summation convention for repeated
indices.

If $f$ is smooth (differentiable) and  convex, then
$\dot{\varepsilon }_{ij}^{P} $ will be unique in any plastic
stress state $\sigma _{ij} $. Under these conditions, the function
$f=\Phi $ can be assigned a flow rule that gives the plastic
strain rate. Within the framework of the plastic potential theory
of Von Mises (1928), each plastic potential $\Phi $ has a flow
rule that is associated with the plastic potential. This flow rule
is given by

\begin{equation} \label{GrindEQ__8_}
\dot{\varepsilon }_{ij}^{P} =\dot{\lambda }\frac{\partial \Phi }{\partial \sigma _{ij} }
\end{equation}

\noindent where $\sigma _{ij} $  is the Cauchy tensor,
$\dot{\varepsilon }_{ij}^{P} $ the plastic strain rate tensor or
plastic flow tensor and $\dot{\lambda }$ a positive plastic
multiplying factor. Based on eq. \eqref{GrindEQ__8_}, if $f=\Phi
$, then the plastic strain rate $\dot{\varepsilon }_{ij}^{P} $ at
the very start of plasticity will be normal to the criterion
surface $f=\Phi $; this constitutes the so-called \textit{plastic
flow normality rule}. Hecker [13] conducted a systematic study of
a large amount of experimental data in metals and found that the
normality rule was never broken. However, there is evidence that
flow normality rule does not hold in soils, granular materials,
etc (materials with non-associated flow rules, where $\partial
f/\partial \sigma _{ij} $ is not proportional to $\partial \Phi
/\partial \sigma _{ij} $) [8, 26].

Defining rate-independent plasticity  requires introducing the
concept of plastic dissipation, which is expressed as

\begin{equation} \label{GrindEQ__9_}
d=\sigma _{ij} \dot{\varepsilon }_{ij}^{P}
\end{equation}

\noindent The parameter\textit{ $d$}  is a measure of the power
per volume that is lost (dissipated), usually as heat, through
deformation. The plastic dissipation as defined in eq.
\eqref{GrindEQ__9_} for the flow rule associated to $f$ will be
maximal if, and only if, the function $f=\Phi $ is convex
\eqref{GrindEQ__9_}; this is the so-called \textit{principle of
maximum plastic dissipation }[31, 16].

Establishing the total dissipation  over a given time interval
entails defining in terms of time-dependent plasticity the amount
of irreversible work per unit volume due to stress as follows:

\begin{equation} \label{GrindEQ__10_}
\chi (t)=W^{P} (t)=\int _{0}^{t}\sigma _{ij} (t)\dot{\varepsilon }_{ij}^{P} (t)dt
\end{equation}

\noindent $\chi (t)$ is also  known as the \textit{work-hardening
parameter }(an internal variable), which is a scalar quantity and
dependent on the plastic time-evolution of the particular solid.

For the study of the hardening  and softening of the solids
materials is necessary the evolution of plasticity function with
the internal variables, $\xi =\xi _{\alpha } $, given by [26]

\noindent

\begin{equation} \label{GrindEQ__11_}
\left. \dot{f}\right|_{\sigma =const.,\; T=const.}  =\sum _{\alpha
}\frac{\partial f}{\partial \xi _{\alpha } }  \dot{\xi }_{\alpha }
=\lambda \sum _{\alpha }\frac{\partial f}{\partial \xi _{\alpha }
}  h_{\alpha } \equiv -\lambda H
\end{equation}

\noindent where $H>0$ for hardening materials,  $H<0$ for
softening materials, and $H=0$ for perfectly plastic materials (in
the latter case $f$ is independent of the $\xi _{\alpha } $). Here
$\lambda $ is a positive continuous function of state variables.

Based on the foregoing, for the theory to be  properly addressed,
the plasticity function $f$ should be convex in the stress-related
variables ($f=\Phi $). The dependency of $f$ on $\xi $ describes
the anisotropy (structural and induced) of materials, while the
dependency of $\dot{f}$ on $\xi $ describes the
hardening-softening of materials.

\section{Plasticity yield criteria}

The following section describes the most relevant plasticity yield
criteria, with emphasis on their particularities (see [32, 33, 15,
18, 2, 22]).

The \textit{Tresca criterion}  [32] which is among the earliest
plasticity yield criteria, can be expressed as a unique function
of the algebraic invariants $(J_{1} ,J_{2} ,J_{3} )$ of the stress
deviation tensor $J$. Based on it, a solid will become plastic
when it reaches a multiaxial state where the multiaxial tangent
stress equals the critical uniaxial tangent stress, $\sigma
_{P}^{} $. This criterion can be expressed as a completely
differentiable relation

\begin{eqnarray}\label{GrindEQ__12_}
f(\sigma _{ij} ;\sigma _{P} )&= \left[(\sigma _{xx} -\sigma _{yy}
)^{2} +4\sigma _{xy}^{2} -4\sigma _{P}^{2} \right][\left(\sigma
_{yy} -\sigma _{zz} )^{2} +4\sigma _{yz}^{2} -4\sigma _{P}^{2}
\right]\nonumber\\
& \times [\left(\sigma _{zz} -\sigma _{xx} )^{2} +4\sigma
_{xz}^{2} -4\sigma _{P}^{2}\right]
\end{eqnarray}

\noindent based on which plasticity  can only be reached on
independent planes, as revealed by factoring the total (volume)
multiaxial state into its partial (surface) multiaxial state. In
other words, plasticity develops on planes. This criterion
exhibits good agreement with experimental results for certain
ductile metals. At microscopic level the movement of dislocations
along slip planes is responsible for permanent deformation [4].
Note that $\sigma _{P}^{} $ depends on the internal variables,
$\xi $, \textit{i.e.} $\sigma _{P}^{} =\sigma _{P}^{} (\xi
)$.\textit{}

The \textit{Huber--Von Mises criterion}  [33, 21] is among the
most widely rule used in this context. It is also known as the
\textit{$J_{2} $}-\textit{criterion} since it is formulated as a
unique function of the algebraic invariant \textit{$J_{2} $ }of
the stress deviation tensor. Based on existing experimental
evidence, this criterion is applicable to ductile materials.
Hencky [14] provided an energy-based interpretation by which the
critical plasticity state is reached when the distortion energy
per volume (\textit{i.e.}, the deformation energy per volume in
the absence of volume changes) in the multiaxial tangent stress
state equals the distortion energy per volume in the critical
uniaxial normal stress state. This criterion is isotropic and can
be formulated as follows:

\begin{equation} \label{GrindEQ__13_}
f(\sigma {}_{ij} ;\sigma _{P} )=(\sigma _{xx} -\sigma _{yy} )^{2}
+(\sigma _{yy} -\sigma _{zz} )^{2} +(\sigma _{zz} -\sigma _{xx}
)^{2} +6(\sigma _{xy}^{2} +\sigma _{yz}^{2} +\sigma _{xz}^{2}
)-2\sigma _{P}^{2}
\end{equation}

\noindent Stress-wise, the Von Mises criterion  indicates that
critical plasticity is reached when the modulus of the multiaxial
tangent stress equals that of the critical uniaxial normal stress,
$\sigma _{P}^{} $. Note that $\sigma _{P}^{} $ depend on the
internal variables, $\xi $, \textit{i.e.} $\sigma _{P}^{} =\sigma
_{P}^{} (\xi )$. Unlike the Tresca criterion, the Von Mises
criterion does not factor the plasticity function into multiaxial
plane stress components; rather, it assumes a mutual dependence
among the multiaxial stresses, which leads to a volume plasticity
expression. In ductile materials, which meet the Von Mises
criterion quite well, plasticity results in distortion by effect
of flow processes occurring with virtually no volume change.

The\textit{ Hill's anisotropic criterion }or  \textit{Hill's first
criterion }[15, 17] provides a general description of materials
with anisotropy (whether structural or induced) and orthotropic
symmetry (\textit{i.e.}, materials where each point possess three
mutually normal planes). Because each plane in an orthotropic
material can be defined in terms of only two parameters, Hill's
criterion can be formulated in terms of six independent
parameters. This constitutes a generalization of the Von Mises
criterion to anisotropic materials of the form

\begin{equation} \label{GrindEQ__14_}
A(\sigma _{yy} -\sigma _{zz} )^{2} +B(\sigma _{zz}  -\sigma _{xx}
)^{2} +C(\sigma _{xx} -\sigma _{yy} )^{2} +2D\sigma _{yz}^{2}
+2E\sigma _{xz}^{2} +2F\sigma _{xy}^{2} =1
\end{equation}

\noindent where $A=A(\xi ),B=B(\xi ),C=C(\xi ),D=D(\xi ),E=E(\xi
),F=F(\xi )$,  depending on the internal variables, are constants
defining the degree of anisotropy in each direction and can be
expressed in terms of the Lankford coefficient. In addition, the
directions $x,y,z$ should be the principal anisotropy directions
for the material ---otherwise, the Cauchy tensor should be
transformed as required in order to have it coincide with the
principal directions. In fact, eq. \eqref{GrindEQ__14_} is a
particular case [26] of

\noindent

\begin{equation} \label{GrindEQ__15_}
\sigma _{ij} B_{ijkl} \sigma _{kl} =\sigma _{P}^{2}
\end{equation}

\noindent which is the more general quadratic form for  the
components of the stress tensor as defined in terms of the
4-tensor $B_{ijkl} $. This tensor fulfils the symmetry conditions
for 4-tensors in the theory of elasticity, which are given by

\noindent

\begin{equation} \label{GrindEQ__16_}
\begin{array}{l} {B_{ijkl} =B_{klij} } \\ {B_{ijkl} =B_{jikl} } \end{array}
\end{equation}

\noindent where the first relation defines the symmetry by  joint
exchange of index pairs $ij\leftrightarrow kl$ and the second the
symmetry by exchange between index pairs of the type
$i\leftrightarrow j$ and/or $k\leftrightarrow l$. This criterion
considers no effects of the mean stress (hydrostatic pressure) on
plasticity; therefore, one must introduce the additional condition
$B_{ijkk} =0$, which allows the quadratic form of such effects to
be neglected.

The \textit{Hosford criterion} [20] and  \textit{Logan--Hosford
criterion} [25] are two generalizations of the Von Mises criterion
that ensure convexity by introducing a parameter \textit{$m\ge
1$}. The Hosford criterion, which is applicable to isotropic
materials, is defined as

\begin{equation} \label{GrindEQ__17_}
f(\sigma _{ij} ;\sigma _{P} )=\frac{1}{2} \left|\sigma _{1}
-\sigma _{2} \right|^{m} +\frac{1}{2} \left|\sigma _{2} -\sigma
_{3} \right|^{m} +\frac{1}{2} \left|\sigma _{1} -\sigma _{3}
\right|^{m} -\sigma _{P}^{m}
\end{equation}

\noindent where $\sigma _{1} ,\sigma _{2} ,\sigma _{3} $ are  the
principal stresses of the Cauchy tensor. Note that $\sigma _{P}^{}
$ depends on the internal variables,$\sigma _{P}^{} =\sigma
_{P}^{} (\xi )$. The Logan--Hosford criterion is a generalization
of Hill's anisotropic criterion \eqref{GrindEQ__14_} that is used
to describe anisotropic materials and expressed as

\noindent

\begin{equation} \label{GrindEQ__18_}
A{'} \left|\sigma _{1} -\sigma _{2} \right|^{m} +B{'}
\left|\sigma _{2} -\sigma _{3} \right|^{m} +C{'} \left|\sigma _{1}
-\sigma _{3} \right|^{m} =1
\end{equation}
where the constants $A{'} =A{'} (\xi ),B{'} =B{'} (\xi ), C{'}
=C{'} (\xi )$, depending on the internal variables, constitute
measures of anisotropy in each direction and can be expressed as a
function of the Lankford coefficient.

\textit{Hill's generalized anisotropic criterion }or
\textit{Hill's second criterion} [18] provides an accurate
description of the anomalous behaviour of some metals such as
aluminium [37]. The criterion is expressed in terms of principal
stresses of the Cauchy tensor:

\begin{eqnarray} \label{GrindEQ__19_}
a\left|\sigma _{2} -\sigma _{3} \right|^{m} +b\left|\sigma _{3}
-\sigma _{1} \right|^{m} +c\left|\sigma _{1} -\sigma _{2}
\right|^{m} +d\left|2\sigma _{1} -\sigma _{2} -\sigma _{3}
\right|^{m} +\nonumber\\
e\left|2\sigma _{2} -\sigma _{1} -\sigma _{3} \right|^{m}
+f\left|2\sigma _{3} -\sigma _{1} -\sigma _{2} \right|^{m} =\sigma
_{P}^{m}
\end{eqnarray}

\noindent where $m\in \Re $ is a parameter  that must fulfil the
condition $m\ge 1$ for the criterion surface to be convex. Also
the constants $a=a(\xi ),b=b(\xi ),c=c(\xi ),d=d(\xi ),e=e(\xi
),f=f(\xi ),\; \sigma _{P}^{} =\sigma _{P}^{} (\xi )$ depending on
the internal variables, $\xi $, and which can be expressed as a
function of the Lankford coefficient, measure anisotropy in each
principal direction. For example, if the principal directions 1--2
in a material define a symmetry plane with mutual isotropy and are
anisotropic with respect to direction 3, then $a=b$ and
\textit{$d=e$} (planar isotropy). On the other hand, if all three
directions are isotropic, then $a=b=c$ and $d=e=f$ (complete
isotropy).

In the \textit{plastic potential theory  of Von Mises} [34], each
plastic potential $\Phi $ is assigned a flow rule. The plastic
potential for the particular case of the Von Mises criterion is
$J_{2} $ and its associated flow rule defined by the Levy--Mises
equations. Koiter [23] developed a generalization of the previous
theory where the plasticity function is defined by a series of
plastic potentials $\Phi _{i} $ each having an associated flow
rule of the type described by eq. \eqref{GrindEQ__8_} above. If
the plasticity function $f$ is expressed as a linear combination
with positive coefficients of the potential functions $\Phi _{i} $
(convex), then $f$ will be convex.

The \textit{plastic anisotropy description}  of Barlat \textit{et
al.} [2] originated from an isotropic plasticity function.
Structural anisotropy in the material was introduced via a series
of linear transformations represented by a 4-tensor acting on the
Cauchy tensor, the latter itself acting on the anisotropic
material. By effect of the transformations, the stress tensor
absorbs structural anisotropy of the material.

The \textit{Karafillis--Boyce criterion}  [22] uses a convex
combination of two plastic potentials as plasticity function. The
potentials are based on Hosford's isotropic criterion
\eqref{GrindEQ__17_} and their degree of mixing is adjusted via
parameter $c\in [0,1]$. Thus, the isotropic criterion is
formulated as

\begin{equation} \label{GrindEQ__20_}
(1-c)\left[\left|s_{1} -s_{2} \right|^{m}  +\left|s_{2} -s_{3}
\right|^{m} +\left|s_{3} -s_{1} \right|^{m} \right]+c\frac{3^{m}
}{2^{m-1} +1} \left[\left|s_{1} \right|^{m} +\left|s_{2}
\right|^{m} +\left|s_{3} \right|^{m} \right]=2\sigma _{P}^{m}
\end{equation}

\noindent where $m$ is positive  and non-zero, and $s_{1} ,s_{2}
,s_{3} $ are the principal values (eigenvalues) of the stress
deviation tensor. Note that $\sigma _{P}^{} ,\; c$ depend on the
internal variables, $\sigma _{P}^{} =\sigma _{P}^{} (\xi ),\; \;
c=c(\xi )$. This criterion is a particular case of Hill's second
criterion \eqref{GrindEQ__19_}. The plasticity yield criterion
\eqref{GrindEQ__20_} represents an isotropic, convex criterion
that can be made anisotropic by applying linear transformations
representing a 4-tensor acting on the Cauchy 2-tensor, which in
turn act on the anisotropic material (Barlat \textit{et al.},
1991). Introducing appropriate symmetries of a material in the
transformation 4-tensor allows all possible states of structural
anisotropy in the material to be considered. The Barlat and
Karafillis--Boyce criteria rely on the \textit{theory of
representation of tensor functions} [35].

Those criteria that consider \textit{ hydrostatic pressure
dependence} assume plasticity in some materials including metallic
foams and polymers to be a function of the hydrostatic pressure
$\sigma _{M} $ acting on the solid. This effect has been
considered by using various general criteria such as those of
\textit{Drucker--Prager}\textbf{\textit{ }}[10,
11],\textbf{\textit{ }}\textit{Caddell}\textbf{\textit{ }}[7] and
\textit{Deshpande }[9], which are modifications of the Von Mises
and Hill criteria including a certain dependence on the
hydrostatic pressure ($\sigma _{M} $).\textit{}

The plasticity yield criterion is  established from the set of
microscopic and macroscopic properties of the material. As a
result, the formulation of each criterion depends on a combination
of parameters of the material describing its anisotropy, crystal
structure and hydrostatic pressure-dependence, among other
properties. There are three general types of models for analysing
plasticity, namely: microstructure or macrostructure and mixed.
\textit{Microstructure models} establish plasticity yield criteria
from the microstructure of each material. On the other hand,
\textit{macrostructure models}, also referred to as
\textit{phenomenological models}, rely on a phenomenological
analysis of the macroscopic behaviour of each material to
establish such criteria. Finally, \textit{mixed models} are
combinations of the previous two and usually provide the more
accurate descriptions of plasticity.

\section{A unified theory of plasticity}

The objective of this paper is to develop a new macroscopic theory
that unifies the plasticity yield criteria. This theory is based
on postulates well-established from experimental data and
theoretical considerations. The used method is based on orthogonal
Lie Groups to describe the classical isotropic yield criteria; an
increase of the symmetry group allows to consider classical and
new anisotropy yield criteria in the solid materials (new
mechanical properties).

At microscopic level, the  plasticity in a solid is produced by a
set of slips, generated by the movement of dislocations, defects,
etc [19]. The stress acting on the solid is the generator of these
slips. We want to take into account all these microscopic
dynamical processes at the macroscopic scale. So the theory is
based on the decomposition of a solid material into parts as
macroscopic points \textit{p} (Figure 1). These macroscopic points
do not need to have any physical reality (grains or others), but
points have macroscopic information about microscopic dynamical
processes. This macroscopic information is purely statistical, so
a macroscopic point is a statistical concept that takes into
account all the microscopic information inside it. The set of
slips inside a macroscopic point induces the irreversible
macroscopic movement of this point (with a well determined spatial
directionality). The global movements of the points result in
macroscopic plasticity that follow macroscopic laws (theory
postulates). So, the proposed unified theory examines plasticity
at each macroscopic point \textit{p} in a solid. The plasticity of
the macroscopic points are described by orthogonal transformation
groups (directionality of plasticity) acting on Cauchy tensor
(generator of plasticity) in a convex manner (measure of
plasticity).

The macroscopic anisotropy  description by coordinate tensors
appears in the infinitesimal orthogonal transformation of Cauchy
stress tensor (which acts on the solid point) with respect to a
reference system (for the solid point). The directions of these
infinitesimal transformations are limited only by the internal
structure (microscopic dynamic processes) of macroscopic solid
points. These directions of transformation (anisotropy
characterization) are described by the coordinate tensors
(parameters that depend on the internal variables). A measure
(mathematical norm) of the infinitesimal orthogonal transformation
of Cauchy tensor gives the macroscopic effect of plasticity
(plasticity function). The ensuing point information can be used
to obtain a global points statistic for the whole solid.

Certain approaches are considered  in this theory. The potential
effects of temperature and strain are ignored
(temperature-independent and rate-independent plasticity), and so
is the time-dependence of all parameters and variables since the
only target is critical plasticity in the solid (time-independent
plasticity). Our theory relies on the following three postulates:

\begin{enumerate}
\item  \textbf{\textit{First postulate:}} \textit{ the macroscopic
units or points that become plastic in a solid are described by
second-order Cauchy tensors.}

\item \textit{ }\textbf{\textit{Second postulate:}}\textit{
hydrostatic-pressure independent plasticity} \textit{acts at every
point in the Tangent Space to the Cauchy Tensor (TSCT). Any type
of plasticity at least needs a component in this space TSCT. }

\item \textit{ }\textbf{\textit{Third postulate:}}\textit{  the
plasticity function is convex in the stress space.}
\end{enumerate}

\noindent Within the framework of the proposed theory,  solids are
represented in a three-dimensional Euclidean space
$E\eqref{GrindEQ__3_}$ where each point or neighbourhood $p\in
E\eqref{GrindEQ__3_}$ is defined by the pair \textit{$p\equiv \{
p^{\lambda } ,\sigma ^{kl} \} \in E\eqref{GrindEQ__3_}$}, with
indices $\lambda ,k,l=x,y,z$. The set of points is referred to
$E\eqref{GrindEQ__3_}$ via a coordinate system or reference system
$\{ O,x^{\lambda } =(x,y,z)\} $ that is shared by the whole solid;
the coordinate system has $O$ as its origin and $x^{\lambda }
=(x,y,z)$ as its axes. Therefore, a point \textit{p} in the solid
lies at $p^{\lambda } $ in the reference system and has the
symmetric second-order stress tensor or Cauchy tensor acting on
it, \textit{i.e.}, $\sigma \equiv \sigma ^{kl} $ (\textit{first
postulate}); this postulate is valid only for solids with
``near-action'' internal forces [24]. Since the solid is examined
point-wise, each point \textit{p} possesses its own reference
system, $\{ O_{p} ,x^{\lambda } =\left({\rm x,\; y,\; z}\right)\}
$, with its origin at $p^{\lambda } $ and $x^{\lambda }
=\left({\rm x,\; y,\; z}\right)$ as axes. So the set of stresses
acting on the points in the solid define a 2-tensor stress field
in the space $E\eqref{GrindEQ__3_}$, which endows the solid with
geometric structure (Figure 1).

The reference system for each point,  $\{ O_{p} ,x^{\lambda }
=\left({\rm x,\; y,\; z}\right)\} $, is arbitrary only when the
ensuing equations are invariant under a reference system change
(tensor equations), that is, independent of the particular
reference system. From the geometric point of view this guarantees
that the tensor equations that describe the solid contain,
codified as geometric information, all the physical behaviour
throughout the directions of reference system changes for the
corresponding space.

If the Cauchy tensor is invariant under  a change in the origin
and axes over the solid, $\{ O_{p_{1} } ,x^{\lambda } =(x,y,z)\}
\to \{ O_{p_{2} } ,\bar{x}^{\lambda } ={\rm
(}\bar{x},\bar{y},\bar{z}{\rm )}\} $, then the solid is
stress-homogeneous. If the Cauchy tensor is invariant under a
change in the axes directions, $\{ O_{p} ,x^{\lambda } =(x,y,z)\}
\to \{ O_{p} ,\bar{x}^{\lambda } ={\rm
(}\bar{x},\bar{y},\bar{z}{\rm )}\} $, then the stresses at the
point \textit{p} in the solid are isotropic. The degree of
homogeneity in the solid can be assessed by examining it
point-wise and using the results to develop a global statistics
for the entire solid. In this work, we will assume the behaviour
of the solid is similar to its individual points or, in other
words, that examining a single point will be the same as studying
the entire solid.

The intrinsic properties of the solid at  each point are examined
via a series of transformations with a fixed origin on it. A
transformation of the Cauchy tensor, $\sigma ^{kl} $, at point
\textit{p} can be interpreted in two completely equivalent ways.
In the first one, the reference system ${\rm x}^{\lambda } $ has
the Cauchy tensor $\sigma ^{kl} $; therefore, a transformation
${\rm x}^{\lambda } \to \bar{{\rm x}}^{\lambda } $ in the
reference system will lead to a new reference system $\bar{{\rm
x}}^{\lambda } $ with a Cauchy tensor $\bar{\sigma }^{ij} $; this
constitutes a reference system transformation or active
transformation. In the other interpretation, the reference system
${\rm x}^{\lambda } $ remains unchanged and the Cauchy tensor
transformation, $\sigma ^{kl} \to \bar{\sigma }^{ij} $, produces a
new tensor that is expressed (oriented) in the reference system
$\bar{{\rm x}}^{\lambda } $; this is a Cauchy tensor
transformation or passive transformation. These two
interpretations are mutually related by a negative sign.\textbf{}

The proposed theory relies on a continuous  linear transformation
of the stress at each point \textit{p} represented by the tensor
$A^{ij} _{kl} $. Such a transformation does not alter the origin
of \textit{p}, which is defined by the coordinates $p^{\lambda }
$, but only its coordinate axes, $x^{\lambda } $. Thus, the
transformation operator $A^{ij} _{kl} $ acts on the stress tensor
$\sigma ^{kl} $ to give the transformed tensor $\bar{\sigma }^{ij}
$ in accordance with the following expression:

\begin{equation} \label{GrindEQ__21_}
A^{ij} _{kl} \sigma ^{kl} =\bar{\sigma }^{ij}
\end{equation}

\noindent where the multilinear transformation  $A^{ij} _{kl} $ is
a 4-tensor of indices $i,j,k,l=x,y,z$ with two covariant $(k,l)$
and two contravariant$(i,j)$ indices. Mathematically, it follows
that this expression is a tensor equation (\textit{i.e.}, that it
is invariant under a coordinate system change). Since the
resulting transformed tensor, $\bar{\sigma }^{ij} $, has a fixed
origin (the same that $\sigma ^{kl} $), the transformation can be
tangent or normal to the tensor $\sigma ^{kl} $. In the former
case, it will produce rotations and/or reflections of use for
studying anisotropy; in the latter case, it will result in
dilatation or dilatation--reflection useful with a view to
examining hydrostatic pressure-dependence.

The plasticity at a point \textit{p} under the  influence of an
arbitrary initial Cauchy tensor depends on the tangent and normal
spaces to the Cauchy tensor. These spaces arise directly from the
geometric generalization of the two types of components of the
Cauchy tensor. The direct sum of the tangent and normal spaces
constitutes the overall stress space at the point concerned. As
implied by our \textit{second postulate}, a solid will yield
plasticity at a given point only if a component on the tangent
space exists at such a point. As a result, plasticity starts in
the associated pure tangent space (hydrostatic-pressure
independent plasticity) or, in other words, it never starts in the
associated pure normal (hydrostatic-pressure dependent plasticity)
space. In fact, there is solid experimental evidence that
plasticity never arises under conditions of pure hydrostatic
pressure [4, 26]. Thus, some solids exhibit slight elastic
(reversible) deformation rather than plasticity, even at very high
pure compressive hydrostatic pressures. However, high
stress-induced hydrostatic pressures in a solid under tension can
lead to an unexpected spatial stress concentration eventually
leading to fragile fracture [5, 6].

\section{Rotational transformation of the Cauchy tensor}

Properly examining plasticity  in a solid entails starting in the
Tangent Space to the Cauchy Tensor (TSCT) associated to each point
\textit{p} in the solid, and applying a pure orthogonal
transformation $R^{ij} _{kl} $ such that

\noindent

\begin{equation} \label{GrindEQ__22_}
R^{ij} _{kl} \sigma ^{kl} =\bar{\sigma }^{ij} ,
\end{equation}

\noindent where the orthogonality condition  is described by means
of the matrix equation $R^{T} R=I$. Based on this tensor equation,
the Cauchy tensor $\sigma ^{kl} $ associated to each point
\textit{p} in the solid is rotated by $R^{ij} _{kl} $ according to
a given parameter, which provides the new associated tensor
$\bar{\sigma }^{ij} $, which will be the new associated tensor
acting on \textit{p}. Stress rotations according to equation
\eqref{GrindEQ__22_} provide a point-wise definition of
hydrostatic-pressure independent plasticity in the solid. In this
sense, rotations of the Cauchy tensor at \textit{p} can be
interpreted physically as due to some tangential stresses acting
on the point. In this respect, the hydrostatic-pressure
independent plasticity interaction (\textit{i.e.}, that having no
hydrostatic component) is fully defined by the rotations $R^{ij}
_{kl} $ and their associated transformation parameters.

The orthogonal rotation transformation $R^{ij} _{kl} $  is a
three-dimensional 4-tensor (\textit{i.e}., one with $3^{4} =81$
components) equivalent to a Mohr transformation if parameterized
in Euler angles $\Theta _{E} =\Theta _{E} (\alpha ,\beta ,\gamma
)$ [27]. As noted earlier, the orthogonal rotation transformation
is necessary for hydrostatic-pressure independent plasticity to
develop at individual points in a solid. However, additional,
non-orthogonal transformations (\textit{e.g.}, dilations) can also
be applied that will alter the hydrostatic-pressure independent
plasticity conditions imposed by a rotation (see Section 11). The
remainder of this section, and all subsequent ones up to the
eleventh, is devoted to examine hydrostatic-pressure independent
plasticity as defined in eq. \eqref{GrindEQ__22_}. Section 11 is
concerned with hydrostatic pressure-dependence.

The transformation equation \eqref{GrindEQ__22_} connects
geometric objects via double indices (\textit{i.e.}, 9
components). However, an equivalent representation connecting
objects via single indices (\textit{i.e.}, 3 components) can be
obtained by using the following relations:

\begin{equation} \label{GrindEQ__23_}
\bar{u}^{i} =R^{i} _{j} u^{j} ;\quad \quad u^{j} =R^{j} _{i} \bar{u}^{i}
\end{equation}

\begin{equation} \label{GrindEQ__24_}
u^{j} =T^{jq} v_{q} ;\quad \quad \bar{u}^{j} =T^{jq} \bar{v}_{q}
\end{equation}

\noindent where \eqref{GrindEQ__23_} defines the rotational
transformations of contravariant vectors and \eqref{GrindEQ__24_}
the general contravariant transformations for a vector under
coordinate system changes (\textit{i.e.}, the definition of a
1-tensor). From the rotational transformation \eqref{GrindEQ__23_}
and general coordinate transformation \eqref{GrindEQ__24_} for
each 1-tensor, one can obtain

\noindent

\begin{equation} \label{GrindEQ__25_}
\left. \begin{array}{l} {\bar{u}^{i} =R^{i} _{j} u^{j} = R^{i}
_{j} T^{jq} v_{q} =R^{i} _{j} T^{jq} R_{q} ^{p} \bar{v}_{p} } \\
{\bar{u}^{i} =\bar{T}^{ip} \bar{v}_{p} }
\end{array}\right\}\Rightarrow \bar{T}^{ip} =R^{i} _{j} T^{jq}
R_{q} ^{p}
\end{equation}

\noindent Since the coordinate system change occurs  via the
stress 2-tensor (Cauchy tensor) $T^{jq} \equiv \sigma ^{jq} $, we
can express the transformation by means of a rotational operator

\begin{equation} \label{GrindEQ__26_}
\bar{\sigma }^{ij} =R^{i} _{k} \sigma ^{kl} R_{l} ^{j}
\Leftrightarrow \bar{\sigma }=R^{T} \sigma R,
\end{equation}

\noindent the latter describing the transformation  rule for the
Cauchy tensor. It is straightforward to verify that this
expression is equivalent to \eqref{GrindEQ__22_}. The 2-tensor
rotations of the types $R^{i} _{k} $ and $R{}_{l} ^{j} $can be
adequately realized in terms of 3 $\times$ 3 matrices. Relation
\eqref{GrindEQ__26_} is invariant under a general change of the
coordinate system.

Therefore, the transformations$R^{i} _{k} $, $R{}_{l} ^{j} $  lead
the plastic process at each point \textit{p} in the solid. Their
study is facilitated by considering the three-dimensional special
orthogonal group \textbf{$SO\eqref{GrindEQ__3_}$ }of rotations in
the three-dimensional Euclidean space $E\eqref{GrindEQ__3_}$. The
main advantage of this approach is that
\textbf{$SO\eqref{GrindEQ__3_}$ }possesses both the structure of a
group of transformations and a differentiable manifold, which
enables us to confer additional invariance properties to the
tensor operators defined over it [12].

The elements of the special orthogonal group
\textbf{$SO\eqref{GrindEQ__3_}$} are the orthogonal
transformations of the Euclidean space $E\eqref{GrindEQ__3_}$ that
describe orientation and length preserving movements. Such
rotations are usually represented by orthogonal real 3 $\times$ 3
matrices $R^{T} R=I$ with unit determinant \textit{$(\det R=+1)$,
}and the group product operation is the usual matrix
multiplication. Topologically, \textbf{$SO\eqref{GrindEQ__3_}$} is
a compact, non-simply connected group [30]. As Lie group,
\textbf{$SO\eqref{GrindEQ__3_}$} is  simple, \textit{i.e.}, the
only normal subgroups it contains are the trivial ones: itself and
the identity group. In particular, its Lie algebra
$so\eqref{GrindEQ__3_}$, which coincides with the tangent space at
the identity element, is also a simple Lie algebra. As a
consequence of the Lie structure the tangent bundle
$TSO\eqref{GrindEQ__3_}\equiv \bigcup _{\rho ^{\nu } \in
SO\eqref{GrindEQ__3_}}TSO\eqref{GrindEQ__3_}_{\rho ^{\nu } }  $
inherits special properties that will be useful in our later
analysis.

\section{Action of the group
\textit{SO\eqref{GrindEQ__3_}} on the Cauchy tensor}

Using the adjoint representation of
\textbf{$SO\eqref{GrindEQ__3_}$}, the action of the rotations
$R^{ij} _{kl} $ in \eqref{GrindEQ__22_} or the equivalent pair
$R^{i} _{k} $, $R{}_{l} ^{j} $ in \eqref{GrindEQ__26_} can be
explicitly expressed in terms of the Euler angles $\Theta _{E}
=\Theta _{E} (\alpha ,\beta ,\gamma )$ as:

\noindent

\begin{equation} \label{GrindEQ__27_}
\left\{\begin{array}{l} {\bar{\sigma }^{xx}  =\sigma ^{xx} \cos
^{2} \gamma +\sigma ^{yy} sin^{2} \gamma +2\sigma ^{xy} sin\gamma
\cos \gamma } \\ {\bar{\sigma }^{yy} =\sigma ^{yy} \cos ^{2}
\gamma +\sigma ^{xx} sin^{2} \gamma -2\sigma ^{xy} sin\gamma \cos
\gamma } \\ {\bar{\sigma }^{zz} =\sigma ^{zz} } \\ {\bar{\sigma
}^{xy} =\sigma ^{xy} (\cos ^{2} \gamma -sin^{2} \gamma )+(\sigma
^{yy} -\sigma ^{xx} )sin\gamma \cos \gamma } \\ {\bar{\sigma
}^{yz} =\sigma ^{yz} \cos \gamma -\sigma ^{xz} sin\gamma } \\
{\bar{\sigma }^{xz} =\sigma ^{yz} sin\gamma +\sigma ^{xz} \cos
\gamma } \end{array}\right. \begin{array}{l} {} \\ {} \end{array}
\end{equation}

\begin{equation} \label{GrindEQ__28_}
\left\{\begin{array}{l} {\bar{\sigma }^{xx}  =\sigma ^{xx} \cos
^{2} \beta +\sigma ^{zz} sin^{2} \beta +2\sigma ^{xz} sin\beta
\cos \beta } \\ {\bar{\sigma }^{yy} =\sigma ^{yy} } \\
{\bar{\sigma }^{zz} =\sigma ^{zz} \cos ^{2} \beta +\sigma ^{xx}
sin^{2} \beta -2\sigma ^{xz} sin\beta \cos \beta } \\ {\bar{\sigma
}^{xy} =\sigma ^{yz} sin\beta +\sigma ^{xy} \cos \beta } \\
{\bar{\sigma }^{yz} =\sigma ^{yz} \cos \beta -\sigma ^{xy}
sin\beta } \\ {\bar{\sigma }^{xz} =\sigma ^{xz} (\cos ^{2} \beta
-sin^{2} \beta )+(\sigma ^{zz} -\sigma ^{xx} )sin\beta \cos \beta
} \end{array}\right. \begin{array}{l} {} \\ {} \end{array}
\end{equation}

\begin{equation} \label{GrindEQ__29_}
\left\{\begin{array}{l} {\bar{\sigma }^{xx}  =\sigma ^{xx} } \\
{\bar{\sigma }^{yy} =\sigma ^{yy} \cos ^{2} \alpha +\sigma ^{zz}
sin^{2} \alpha -2\sigma ^{yz} sin\alpha \cos \alpha } \\
{\bar{\sigma }^{zz} =\sigma ^{zz} \cos ^{2} \alpha +\sigma ^{yy}
sin^{2} \alpha +2\sigma ^{yz} sin\alpha \cos \alpha } \\
{\bar{\sigma }^{xy} =\sigma ^{xy} \cos \alpha -\sigma ^{xz}
sin\alpha } \\ {\bar{\sigma }^{yz} =\sigma ^{yz} (\cos ^{2} \alpha
-sin^{2} \alpha )+(\sigma ^{yy} -\sigma ^{zz} )sin\alpha \cos
\alpha } \\ {\bar{\sigma }^{xz} =\sigma ^{xy} sin\alpha +\sigma
^{xz} \cos \alpha } \end{array}\right.
\end{equation}

A rotation at the point \textit{p} therefore  represents a
transformation from the reference system ${\rm \{ O}_{{\rm p}}
{\rm ,x}^{\lambda } \} $, with a tensor $\sigma ^{kl} $, to the
new reference system ${\rm \{ O}_{{\rm p}} {\rm ,}\bar{{\rm
x}}^{\lambda } \} $, with associated tensor $\bar{\sigma }^{ij} $.
The components of the Cauchy tensor in the rotated reference
system $\bar{\sigma }^{ij} $ are expressed in terms of $\sigma
^{kl} $ by means of equations \eqref{GrindEQ__27_} -
\eqref{GrindEQ__29_}.

The rotation by angle $\Theta _{E} =\Theta _{E}  (\alpha ,\beta
,\gamma )$ can be easily visualized by viewing the reference
system in the direction from the positive ends of the axes to
their origin and rotating $x^{\lambda } $ anti-clockwise to the
new system, $\bar{x}^{\lambda } $, by an angle \textit{$\alpha$},
\textit{$\beta$} and \textit{$\gamma$} about the \textit{x},
\textit{y} and \textit{z} axis, respectively (see Figures 2 and
3). Obviously, because of the non-Abelianity of
\textbf{$SO\eqref{GrindEQ__3_}$}, the order in which the three
partial rotations are performed strongly influences the general
rotational outcome.

The action of a generic rotation on the Cauchy  tensor is the
result of applying successive partial rotations (Figure 3) about
each axis in a given sequence. Thus, applying the rotations
following the sequence $\alpha \to \beta \to \gamma $, we get

\begin{equation} \label{GrindEQ__30_}
\bar{\sigma }^{ij} =R^{ij} _{kl} (\Theta _{E} )\sigma ^{kl}
=R^{ij} _{kl} (\gamma ,\beta ,\alpha )\sigma ^{kl}
\end{equation}
\textit{}

\noindent which is equivalent to the following  expression for
rotations in the group \textbf{$SO\eqref{GrindEQ__3_}$}:

\noindent

\begin{equation} \label{GrindEQ__31_}
\bar{\sigma }^{ij} =[R(\gamma )R(\beta )R(\alpha )]^{i}_{k} \sigma
^{kl} [R(\alpha )R(\beta )R(\gamma )]_{l} ^{j} \Leftrightarrow
\bar{\sigma }=R^{T} (\gamma ,\beta ,\alpha )\sigma R(\alpha ,\beta
,\gamma )
\end{equation}

\noindent with partial rotations performed following the  sequence
$\alpha \to \beta \to \gamma $. We observe that, although
equations \eqref{GrindEQ__27_} - \eqref{GrindEQ__29_} depend on
the explicit choice of rotation parameters, the transformation
does not alter the form of equations \eqref{GrindEQ__30_} and
\eqref{GrindEQ__31_}, which, as noted earlier, are invariant under
a change of coordinate system.

\noindent

\section{Infinitesimal action of \textit{SO\eqref{GrindEQ__3_}} on the Cauchy tensor}

Because of the properties of Lie algebras and the local
diffeormorphism on a neighbourhood of the identity element [12],
the action \eqref{GrindEQ__31_} on the Cauchy tensor can be also
described in terms of the infinitesimal rotations, which turn out
to be the generators of $so\eqref{GrindEQ__3_}$.

The Lie group \textbf{$SO\eqref{GrindEQ__3_}$ }and  its associated
Lie algebra,\textbf{ }$so\eqref{GrindEQ__3_}$, are related via the
\textit{exponential mapping}. Given an element $X_{\nu } \in
so\eqref{GrindEQ__3_}$, the exponential map provides an element of
\textbf{$SO\eqref{GrindEQ__3_}$ }by means of the relation:

\noindent

\begin{equation} \label{GrindEQ__32_}
Exp(X_{\nu } )=R(\theta ^{\nu } )=e^{\theta ^{\nu } X_{\nu } } .
\end{equation}

\noindent Since \textbf{$SO\eqref{GrindEQ__3_}$ }is a  compact
group, it is ensured that \textit{Exp} is a surjective mapping
[30]. Differentiation with respect to the parameter and evaluation
at the identity element provides the operator $X$:

\noindent

\begin{equation} \label{GrindEQ__33_}
\begin{array}{c} {\left. \frac{\partial }{\partial \theta ^{\nu } }
R(\theta ^{\nu } )\right|_{\theta ^{\nu } =0} =X_{\nu } } \\
{\left. \frac{d}{d\Theta } R(\Theta _{E} )\right|_{\Theta =0}
=\rho ^{\nu } \left. \frac{\partial }{\partial \theta ^{\nu } }
R(\theta ^{\nu } )\right|_{\theta ^{\nu } =0} =\rho ^{\nu } X_{\nu } =X}
\end{array}
\end{equation}

The operators $X$ in the Lie algebra therefore correspond  to
infinitesimal rotations and can be expressed as linear
combinations of the operators for the local basis $X_{\nu } \equiv
\partial /\partial \theta ^{\nu } $ in the Lie algebra
$so\eqref{GrindEQ__3_}$. As a result, each infinitesimal rotation
can be expressed as $X=\rho ^{\nu } X_{\nu } \equiv \rho ^{\nu }
\left(\partial /\partial \theta ^{\nu } \right)$, which has $\rho
^{\nu } $ as its local coordinates.

The actions of infinitesimal rotations on the stress field  are
determined by the exponential map \eqref{GrindEQ__32_}, which
relates the group and the Lie algebra; this equation is a function
of the infinitesimal generators $X_{\nu } $ of each partial
rotation. A first-order series expansion about the angle $\theta
^{\nu } =0$ (group identity) yields the expression:

\begin{equation} \label{GrindEQ__34_}
R(\theta ^{\nu } )=e^{\theta ^{\nu } X_{\nu } } =I+X_{\nu } \delta \theta ^{\nu } +...
\end{equation}

\noindent where $\Theta _{E} =\Theta _{E} (\alpha ,\beta ,\gamma )
\equiv \Theta _{E} (\theta ^{1} ,\theta ^{2} ,\theta ^{3} )$
denote the Euler angles associated to the first-order
infinitesimal rotations $\delta \theta ^{\nu } $, which can take
the values $\delta \theta ^{\nu } =\delta \alpha ,\delta \beta
,\delta \gamma $. In addition, each infinitesimal angle $\delta
\theta ^{\nu } $ has an associated infinitesimal generator $X_{\nu
} \equiv \partial /\partial \theta ^{\nu } $. Substituting
\eqref{GrindEQ__34_} into the rotational expression for a 2-tensor
[eq. \eqref{GrindEQ__31_}] yields the expression describing the
action of partial infinitesimal rotations, by an angle $\theta
^{\nu } $, on the Cauchy tensor. For a first-order approximation,
such an expression is of the form

\noindent

\begin{equation} \label{GrindEQ__35_}
\begin{array}{l} {\sigma (\theta ^{\nu } +\delta \theta ^{\nu } )
=\left(I+X_{\nu } \delta \theta ^{\nu } \right)^{T} \sigma (\theta ^{\nu } )
\left(I+X_{\nu } \delta \theta ^{\nu } \right)=} \\ {\; \; \; \; \; \; \quad
 \; \quad \quad \; =\sigma (\theta ^{\nu } )+\{ \sigma (\theta ^{\nu } )
 X_{\nu } +X_{\nu } ^{T} \sigma (\theta ^{\nu } )\} \delta \theta ^{\nu } +
 O^{2} \left(\delta \theta ^{\nu } \right)} \end{array}
\end{equation}

\noindent where eq. \eqref{GrindEQ__31_} has been applied on the
assumption that $\bar{\sigma }=\sigma (\theta ^{\nu } +\delta
\theta ^{\nu } )$ and $\sigma =\sigma (\theta ^{\nu } )$
(\textit{i.e.}, the neither the origin nor the rotation centre of
the tensor is altered, since the transformation of the Cauchy
tensor involves rotations only, so that the tensor remains
unchanged but is evaluated at a different angle). Equation
\eqref{GrindEQ__35_} can be used to establish the first-order
variation at each partial angle, which is of the form

\noindent

\begin{equation} \label{GrindEQ__36_}
\begin{array}{c} {\sigma (\theta ^{\nu } +\delta \theta ^{\nu } )
-\sigma (\theta ^{\nu } )=\sigma (\theta ^{\nu } )X_{\nu } \delta
\theta ^{\nu } +\left(X_{\nu } \right)^{T} \sigma (\theta ^{\nu } )
\delta \theta ^{\nu } \Leftrightarrow } \\ {\Leftrightarrow
\frac{\delta \sigma }{\delta \theta ^{\nu } } =\sigma X_{\nu }
+\left(X_{\nu } \right)^{T} \sigma =[\sigma ,X_{\nu } ]} \end{array}
\end{equation}

\noindent where the infinitesimal generators $X_{\nu } $ for
\textbf{$SO\eqref{GrindEQ__3_}$ }are antisymmetric and the binary
operation\textbf{ }$[\sigma ,X_{\nu } ]=\sigma X_{\nu } -X_{\nu }
\sigma $ is the commutator between the stress field and the
infinitesimal generators $X_{\nu } $. Equation
\eqref{GrindEQ__36_} is a function of the specific basis $X_{\nu }
$ associated to the Euler angles. Expressing it in terms of an
invariant generic infinitesimal rotation $X$ describing the
tangent spaces to \textbf{$SO\eqref{GrindEQ__3_}$ }in any
arbitrary coordinate system entails introducing the local
coordinates\textbf{ }$\rho ^{\nu } $ associated to the local basis
$X_{\nu } $ for the infinitesimal rotation $X$.

This results in a \textit{first-order variation}  of the tensor
stress field $\sigma $ with the invariant generator (infinitesimal
rotation) $X=\rho ^{\nu } X_{\nu } $ of the form

\noindent

\begin{equation} \label{GrindEQ__37_}
\left. d\sigma \right|_{1} \equiv [\sigma ,X]= \sigma X-X\sigma
=\rho ^{\nu } \frac{\delta \sigma }{\delta \theta ^{\nu } }
\end{equation}
which is the equation for a 2-tensor. Since the operator  $X$ is
antisymmetric, then $\left. d\sigma \right|_{1} \equiv [\sigma
,X]=-[X,\sigma ]\equiv -\left. dX\right|_{1} $, based on which the
variation of $\sigma $ with $X$ is equivalent to that of $X$ with
$\sigma $ with the opposite sign. This is consistent with the two
interpretations of a transformation provided in Section 4 as
applied to a first-order variation.

The tangent space produced by the first-order variation
\eqref{GrindEQ__37_} is a linear combination of Euler angles (or,
similarly, generators $X_{\nu } $) where each term is linearly
dependent on the angles (or generators); therefore, eq.
\eqref{GrindEQ__37_} comprises a linear basis of generators
$X_{\nu } $ (an Euclidean space). This linear approximation
describes TSCT at a first-order local level only (\textit{i.e}.,
in very small or local neighbourhoods) (see Figure 4). Expanding
such neighbourhoods entails using higher-order approximations
involving non-linear combinations in the $X_{\nu } $ generator
basis (\textit{i.e.}, non-Euclidean spaces for the different
orders) (Figure 5). As shown below by explicit calculations, all
spaces approximating TSCT are traceless (\textit{i.e.}, the
complete approximation are in fact tangent to the Cauchy tensor).
This allows TSCT to be described by means of expansion series of
different order.\textbf{}

If one assumes anisotropy at each point in a solid to  be more
accurately described by including higher order-terms in the
series, then the geometric interpretation of such anisotropy
suggests that the more local the space ---and the smaller the
neighbourhood---, the higher will be the degree of isotropy. The
inclusion of terms of an increasingly higher order gradually makes
the local structure complex enough to facilitate extension of each
infinitesimal neighbourhood, with improves approximation, to the
entire TSCT. The complex behaviour of solids as concerning
symmetry, anisotropy and various other properties dictates the
structure of TSCT and hence the approximation orders needed for an
accurate description.

Higher-order variations are calculated by applying eq.
\eqref{GrindEQ__37_} recursively or, alternatively, expanding the
total rotation in \eqref{GrindEQ__31_} up to the required order.

The \textit{second-order variation} of the stress field  $\sigma $
with the rotation $X=\rho ^{\nu } X_{\nu } $ can thus be expressed
as the following symmetric 2-tensor:

\begin{equation} \label{GrindEQ__38_}
\left. d\sigma \right|_{2} \equiv [[\sigma ,X],X]= \sigma
XX-2X\sigma X+XX\sigma =\rho ^{\nu \mu } \frac{\delta ^{2} \sigma
}{\delta \theta ^{\nu } \delta \theta ^{\mu } }
\end{equation}

\noindent which represents the  second-order approximation space
with coordinates $\rho ^{\nu \mu } =\rho ^{\nu } \rho ^{\mu } $ in
the operator basis $X_{\nu \mu } =X_{\nu } X_{\mu }
=\frac{\partial ^{2} }{\partial \theta ^{\nu } \partial \theta
^{\mu } } $, the action of which on $\sigma $ being self-apparent
from this expression.

The \textit{third-order variation}  of the stress field $\sigma $
with the rotation $X=\rho ^{\nu } X_{\nu } $ can be expressed as
the following symmetric 2-tensor:

\begin{equation} \label{GrindEQ__39_}
\left. d\sigma \right|_{3} \equiv [[[\sigma ,X],X],X]=\sigma
XXX-3X\sigma XX+3XX\sigma X-XXX\sigma =\rho ^{\nu \mu \omega }
\frac{\delta ^{3} \sigma }{\delta \theta ^{\nu } \delta \theta
^{\mu } \delta \theta ^{\omega } }
\end{equation}

\noindent This equation represents the third-order  approximation
space with coordinates $\rho ^{\nu \mu \omega } =\rho ^{\nu } \rho
^{\mu } \rho ^{\omega } $ in the operator basis $X_{\nu \mu \omega
} =X_{\nu } X_{\mu } X_{\omega } =\frac{\partial ^{3} }{\partial
\theta ^{\nu } \partial \theta ^{\mu } \partial \theta ^{\omega }
} $, the action of which on $\sigma $ is also straightforward.

Higher order approximations are computed by following  the same
procedure. As a result one obtains a symmetric \textit{n}-tensor
of coordinates $\rho ^{\nu ...\kappa } =\rho ^{\nu } \cdot
...\cdot \rho ^{\kappa } $ in the basis $X_{\nu ...\kappa }
=X_{\nu } \cdot \ldots \cdot X_{\kappa } =\frac{\partial ^{n}
}{\partial \theta ^{\nu } \cdot \ldots \cdot \partial \theta
^{\kappa } } $ for each \textit{n}-order. Also, each
\textit{n}-order variation encompasses a total of
$\left(\begin{array}{c} {3+n-1} \\ {n} \end{array}\right)$ terms,
each of which can be factored into sets of \textit{n} generators
$X_{\nu } $, where the resulting factors represent the basis
$X_{\nu ...\kappa } $ in the \textit{n}-order approximation space.
Therefore, each \textit{n}-order is represented by the
corresponding coordinates $\rho ^{\nu ...\kappa } $ in the basis
$X_{\nu ...\kappa } $, and the coordinates represent a symmetric
\textit{n}-tensor with $\left(\begin{array}{c} {3+n-1} \\ {n}
\end{array}\right)$ independent components. Note that the
\textit{n}-tensors of coordinates depend on the internal
variables, $\rho ^{\nu } =\rho ^{\nu } (\xi ),\; \ldots ,\; \rho
^{\nu ...\kappa } =\rho ^{\nu ...\kappa } (\xi )$; tensors of
coordinates can be properly taken as the \textit{position internal
variables} that describe anisotropy.

Each variation has an equivalent interpretation,  whatever its
order, since transforming the axes of the reference system is
equivalent to transforming the stresses involved ---with a
negative sign (see Section 4) by virtue of the antisymmetric
nature of the Lie bracket.

The infinitesimal generators of \textbf{$SO\eqref{GrindEQ__3_}$}
conform to\textbf{ }the following commutation relations:

\begin{equation} \label{GrindEQ__40_}
[X_{a} ,X_{b} ]=\varepsilon _{abc} X_{c}
\end{equation}

\noindent where $\varepsilon _{abc} $ is the Levi--Civita symbol.
These generators can be expressed as 2-tensors of the form

\noindent

\begin{equation} \label{GrindEQ__41_}
\begin{array}{l} {X_{1} =\frac{\partial }{\partial \alpha }
=\left(\begin{array}{ccc} {0} & {0} & {0} \\ {0} & {0} & {-1} \\
{0} & {1} & {0} \end{array}\right)} \\ {X_{2} =\frac{\partial
}{\partial \beta } =\left(\begin{array}{ccc} {0} & {0} & {1} \\
{0} & {0} & {0} \\ {-1} & {0} & {0} \end{array}\right)} \\ {X_{3}
=\frac{\partial }{\partial \gamma } =\left(\begin{array}{ccc} {0}
& {-1} & {0} \\ {1} & {0} & {0} \\ {0} & {0} & {0}
\end{array}\right)} \end{array}
\end{equation}

In summary, using the three generators of the group
\textbf{$SO\eqref{GrindEQ__3_}$} in eq. \eqref{GrindEQ__41_} in
conjunction with the Cauchy tensor\textbf{ }$\sigma $ allows us to
compute the variation of any order of the Cauchy tensor in order
to obtain the different approximation to TSCT.

\noindent \underbar{}

\section{$\mathbf{n}$-Order variations of the stress tensor field}

This section specifically examines the variations of different
order described in the previous section, which provide different
approximations to TSCT.

The first-order variation \eqref{GrindEQ__37_} of $\sigma $ with
$X$can be expressed as

\begin{equation}%
\begin{array}
[c]{l}%
{\lbrack\sigma,X]=\rho^{1}\left(
\begin{array}
[c]{ccc}%
{0} & {\sigma_{xz}} & {-\sigma_{xy}}\\
{\sigma_{xz}} & {2\sigma_{yz}} & {\sigma_{zz}-\sigma_{yy}}\\
{-\sigma_{xy}} & {\sigma_{zz}-\sigma_{yy}} & {-2\sigma_{yz}}%
\end{array}
\right)  +\rho^{2}\left(
\begin{array}
[c]{ccc}%
{-2\sigma_{xz}} & {-\sigma_{yz}} & {\sigma_{xx}-\sigma_{zz}}\\
{-\sigma_{yz}} & {0} & {\sigma_{xy}}\\
{\sigma_{xx}-\sigma_{zz}} & {\sigma_{xy}} & {2\sigma_{xz}}%
\end{array}
\right)  +}\\
\quad\quad\quad\quad {\rho^{3}\left(
\begin{array}
[c]{ccc}%
{2\sigma_{xy}} & {\sigma_{yy}-\sigma_{xx}} & {\sigma_{yz}}\\
{\sigma_{yy}-\sigma_{xx}} & {-2\sigma_{xy}} & {-\sigma_{xz}}\\
{\sigma_{yz}} & {-\sigma_{xz}} & {0}%
\end{array}
\right)  =}\\
{\left(
\begin{array}
[c]{ccc}%
{-\rho^{2}2\sigma_{xz}+2\rho^{3}\sigma_{xy}} &
{\rho^{1}\sigma_{xz}-\rho
^{2}\sigma_{yz}+\rho^{3}(\sigma_{yy}-\sigma_{xx})} &
{-\rho^{1}\sigma
_{xy}+\rho^{2}(\sigma_{xx}-\sigma_{zz})+\rho^{3}\sigma_{yz}}\\
{\rho^{1}\sigma_{xz}-\rho^{2}\sigma_{yz}+\rho^{3}(\sigma_{yy}-\sigma_{xx})}
&
{\rho^{1}2\sigma_{yz}-\rho^{3}2\sigma_{xy}} & {\rho^{1}(\sigma_{zz}%
-\sigma_{yy})+\rho^{2}\sigma_{xy}-\rho^{3}\sigma_{xz}}\\
{-\rho^{1}\sigma_{xy}+\rho^{2}(\sigma_{xx}-\sigma_{zz})+\rho^{3}\sigma_{yz}}
&
{\rho^{1}(\sigma_{zz}-\sigma_{yy})+\rho^{2}\sigma_{xy}-\rho^{3}\sigma_{xz}}
&
{-\rho^{1}2\sigma_{yz}+\rho^{2}2\sigma_{xz}}%
\end{array}
\right)  }%
\end{array}
\label{GrindEQ__42_}%
\end{equation}

\noindent which is a real symmetric traceless  2-tensor
parameterized in the coordinates $\rho ^{\nu } $. The first-order
variation for the particular case of the Cauchy diagonal tensor is

\noindent

\begin{equation} \label{GrindEQ__43_}
[\sigma ,X]=\left(\begin{array}{ccc} {0} & {\rho ^{3}(\sigma _{2}
-\sigma _{1} )} & {\rho ^{2} (\sigma _{1} -\sigma _{3} )} \\ {\rho
^{3} (\sigma _{2} -\sigma _{1} )} & {0} & {\rho ^{1} (\sigma _{3}
-\sigma _{2} )} \\ {\rho ^{2} (\sigma _{1} -\sigma _{3} )} & {\rho
^{1} (\sigma _{3} -\sigma _{2} )} & {0} \end{array}\right)
\end{equation}

\noindent and the second-order variation \eqref{GrindEQ__38_}  of
$\sigma $ with $X$ is

\noindent

\begin{equation} \label{GrindEQ__44_}
\begin{array}{l} {[[\sigma ,X],X]^{11} =\rho ^{12} 2\sigma _{xy}
+\rho ^{13} 2\sigma _{xz} -\rho ^{23} 4\sigma _{yz} +\rho ^{33}
2(\sigma _{yy} -\sigma _{xx} )+\rho ^{22} 2(\sigma _{zz} -\sigma
_{xx} );}\\ {[[\sigma ,X],X]^{12} =-(\rho ^{11} +\rho ^{22} +4\rho
^{33} )\sigma _{xy} +\rho ^{23} 3\sigma _{xz} +\rho ^{13} 3\sigma
_{yz} +\rho ^{12} (-2\sigma _{zz} +\sigma _{yy} +\sigma _{xx} );}
\\ {[[\sigma ,X],X]^{13} =\rho ^{23} 3\sigma _{xy} -(\rho ^{11}
+\rho ^{33} +4\rho ^{22} )\sigma _{xz} +\rho ^{12} 3\sigma _{yz}
+\rho ^{13} (-2\sigma _{yy} +\sigma _{xx} +\sigma _{zz} );} \\
{[[\sigma ,X],X]^{22} =2\rho ^{12} \sigma _{xy} -\rho ^{13}
4\sigma _{xz} +\rho ^{23} 2\sigma _{yz} -\rho ^{33} 2(\sigma _{yy}
-\sigma _{xx} )+\rho ^{11} 2(\sigma _{zz} -\sigma _{yy} );} \\
{[[\sigma ,X],X]^{23} =\rho ^{13} 3\sigma _{xy} +\rho ^{12}
3\sigma _{xz} +\rho ^{23} (-2\sigma _{xx} +\sigma _{yy} +\sigma
_{zz} )-(4\rho ^{11} +\rho ^{22} +\rho ^{33} )\sigma _{yz} ;} \\
{[[\sigma ,X],X]^{33} =-\rho ^{12} 4\sigma _{xy} +\rho ^{13}
2\sigma _{xz} +\rho ^{23} 2\sigma _{yz} -\rho ^{22} 2(\sigma _{zz}
-\sigma _{xx} )-\rho ^{11} 2(\sigma _{zz} -\sigma _{yy} )}
\end{array}
\end{equation}

\noindent which is a real symmetric traceless 2-tensor
parameterized in the coordinates $\rho ^{\nu \mu } =\rho ^{\nu }
\rho ^{\mu } $. The specific expression for the Cauchy diagonal
tensor is

\noindent

\begin{eqnarray} \label{GrindEQ__45_}
\left[\left[\sigma
,X\right],X\right]=\nonumber\\
\left(\begin{array}{ccc} {\rho ^{33}
2(\sigma _{2} -\sigma _{1} )+\rho ^{22} 2(\sigma _{3} -\sigma _{1}
)} & {\rho ^{12} (-2\sigma _{3} +\sigma _{2} +\sigma _{1} ))} &
{\rho ^{13} (-2\sigma _{2} +\sigma _{1} +\sigma _{3} )}
\\ {\rho ^{12} (-2\sigma _{3} +\sigma _{2} +\sigma _{1} )} &
{-\rho ^{33} 2(\sigma _{2} -\sigma _{1} )+\rho ^{11} 2(\sigma _{3}
-\sigma _{2}
)} & {\rho ^{23} (-2\sigma _{1} +\sigma _{2} +\sigma _{3} )} \\
{\rho ^{13} (-2\sigma _{2} +\sigma _{1} +\sigma _{3} )} & {\rho
^{23} (-2\sigma _{1} +\sigma _{2} +\sigma _{3} )} & {-\rho ^{22}
2(\sigma _{3} -\sigma _{1} )-\rho ^{11} 2(\sigma _{3} -\sigma _{2}
)} \end{array}\right)
\end{eqnarray}

The third-order variation \eqref{GrindEQ__39_} of $\sigma $ with
$X$ for the particular case of the Cauchy diagonal tensor is given
by:

\noindent

\begin{equation} \label{GrindEQ__46_}
\begin{array}{l} {[[[\sigma ,X],X],X]^{11} =
\rho ^{123} 6(\sigma _{2} -\sigma _{3} );} \\
{[[[\sigma ,X],X],X]^{12} =\rho ^{333} 4(\sigma _{1} -\sigma _{2}
)+\rho ^{223} (4\sigma _{1} -3\sigma _{3} -\sigma _{2} )+\rho
^{133} (-4\sigma _{2} +3\sigma _{3} +\sigma _{1} );} \\ {[[[\sigma
,X],X],X]^{13} =\rho ^{233} (-4\sigma _{1} +3\sigma _{2} +\sigma
_{3} )+\rho ^{112} (4\sigma _{3} -3\sigma _{2} -\sigma _{1} )+\rho
^{222} 4(\sigma _{3} -\sigma _{1} );} \\ {[[[\sigma ,X],X],X]^{22}
=\rho ^{123} 6(\sigma _{3} -\sigma _{1} );} \\ {[[[\sigma
,X],X],X]^{23} =\rho ^{133} (4\sigma _{2} -3\sigma _{1} -\sigma
_{3} )+\rho ^{111} 4(\sigma _{2} -\sigma _{3} )+\rho ^{122}
(-4\sigma _{3} +3\sigma _{1} +\sigma _{2} );} \\ {[[[\sigma
,X],X],X]^{33} =\rho ^{123} 6(\sigma _{1} -\sigma _{2} )}
\end{array}
\end{equation}

\noindent which is a real symmetric traceless  2-tensor
parameterized in the coordinates $\rho ^{\nu \mu \omega } =\rho
^{\nu } \rho ^{\mu } \rho ^{\omega } $.

The following series expansion encompasses those for  all
variation orders and defines the overall TSCT:

\begin{equation} \label{GrindEQ__47_}
d\sigma =\left. d\sigma \right|_{1} +\frac{1}{2!} \left. d\sigma
\right|_{2} +\frac{1}{3!} \left. d\sigma \right|_{3} +...=[\sigma
,X]+\frac{1}{2!}  [[\sigma ,X],X]+\frac{1}{3!} [[[\sigma
,X],X],X]+.....
\end{equation}

Therefore, TSCT is represented by a symmetric traceless  2-tensor
\eqref{GrindEQ__47_} that describes hydrostatic-pressure
independent plasticity. If the solid concerned is completely
isotropic, using the first term in \eqref{GrindEQ__47_} will
suffice; if it is orthotropic, the first two will be enough. The
plastic processes independent of the hydrostatic pressure are
completely characterized by the stress dependences and tensors of
coordinates $\rho ^{\nu } ,\ldots ,\rho ^{\nu \ldots \kappa } $
---in their corresponding bases--- in the terms for the different
variation orders in \eqref{GrindEQ__47_}.

\section{The norm space in plastic states}

The geometric  study of plasticity conducted here involves
examining hydrostatic-pressure independent plasticity in the
tangent space to the Cauchy tensor (TSCT). As shown above, such a
space can be represented by the series expansion
\eqref{GrindEQ__47_}, which has the typical form for a 2-tensor
and gives both the spatial coordinates (degree of anisotropy) in a
specific basis and the stress-dependence of hydrostatic-pressure
independent plasticity. However, determining the plasticity
potential (plasticity interaction) in this situation entails
defining a specific metric for the 2-tensor \eqref{GrindEQ__47_}.
To this extend, expression \eqref{GrindEQ__47_} ---or the
corresponding partial summation--- is used to calculate the scalar
norm $g$. The parameters introduced by the definition of the norm
are restricted by the constraint that the plasticity function
should be convex (\textit{third postulate}).

Plasticity depends both on the externally  applied stresses,
$\sigma ^{kl} $, and on the intrinsic properties of the material,
$\rho ^{\nu } ,\ldots ,\rho ^{\nu \ldots \kappa } $, in a specific
reference system or basis. Therefore, the scalar function $g$
defining the norm will depend on both [\textit{i.e.}, $g=g(\sigma
^{kl} ;\rho ^{\nu } ,...,\rho ^{\nu ...\kappa } )$]. The function
$g$ should provide the correlation between multiaxial stresses and
hence indicate how the different components of the Cauchy tensor
must be related. If \textit{$(\sigma ^{kl} )_{P} $} is defined as
the 2-tensor describing the plastic limit of the solid
(\textit{stress internal variables}), then such a limit will be
reached when

\begin{equation} \label{GrindEQ__48_}
g(\sigma ^{kl} ;\rho ^{\nu } ,...,\rho ^{\nu ...\kappa } )
=g((\sigma ^{kl} )_{P} ;\rho ^{\nu } ,...,\rho ^{\nu ...\kappa } )
\end{equation}

Usually, the tensor \textit{$(\sigma ^{kl} )_{P} $} encompasses  a
single non-zero uniaxial component and thus coincides with
\textit{$\sigma _{P} $} (the scalar uniaxial plastic limit that it
is an internal variable). This allows the plasticity function to
be rewritten as follows:

\begin{equation} \label{GrindEQ__49_}
f(\sigma ^{kl} ,(\sigma ^{kl} )_{P} ;\rho ^{\nu } ,...,\rho ^{\nu ...\kappa } )
=g(\sigma ^{kl} ;\rho ^{\nu } ,...,\rho ^{\nu ...\kappa } )-g((\sigma ^{kl} )_{P}
;\rho ^{\nu } ,...,\rho ^{\nu ...\kappa } )
\end{equation}

The scalar function $g$ can be defined in terms of a  matrix norm
$\left\| \cdot \right\| $ such as $L^{m} $ (\textit{i.e.} the
usual \textit{m}-norm for vectors); therefore,

\begin{equation} \label{GrindEQ__50_}
\begin{array}{c} {\left\| d\sigma \right\| ^{m} \equiv
\left(\sum _{i=1}^{3}\sum _{j=1}^{3}\left|d\sigma _{ij}
\right|^{m}   \right)} \\ {g=\left\| d\sigma \right\| ^{m}
\Leftrightarrow f=\left\| d\sigma \right\| ^{m} -\left\|
\left(d\sigma \right)_{P} \right\| ^{m} } \end{array}
\end{equation}

\noindent where the constraints on the norm parameter  ($m\in \Re
,^{} m\ge 1$) ensure convexity in the plasticity yield criterion
(\textit{third postulate}). The specific value of \textit{m}
depends on the properties of the particular material. Norm
\eqref{GrindEQ__50_} can be induced from the scalar product of
second-order tensors, which is constructed from the following:

\noindent

\begin{equation} \label{GrindEQ__51_}
<d\sigma ,\; d\sigma {'} >\equiv d\sigma ^{kl} d\sigma _{kl} {'}
\stackrel{R}{\longrightarrow}<R^{T} d\sigma R,\; R^{T} d\sigma {'}
R>\equiv (R^{i} _{k} d\sigma ^{kl} R_{l} ^{j} )(R_{i} ^{k} d\sigma
{'} _{kl} R^{l} _{j} )=d\sigma ^{kl} d\sigma _{kl} {'}
\end{equation}

\noindent This definition, where rotations $R^{i} _{k} $  and
$R_{l} ^{j} $ are orthogonal \textit{$(R^{T} R=I)$}, is invariant
under rotations (\textbf{\textit{R}}). The scalar product and its
induced norm [eqs \eqref{GrindEQ__50_} and \eqref{GrindEQ__51_}]
afford measurement of angles and lengths in TSCT
\eqref{GrindEQ__47_}.

The definition must be completed with a scalar product of  the
coordinate tensors at the different independent orders ($\rho
^{\nu } ,\ldots ,\rho ^{\nu \ldots \kappa } $) of the form

\begin{equation} \label{GrindEQ__52_}
\rho ^{\nu ...\kappa } \rho _{\chi ....\xi } = \left(\rho ^{\nu }
\right)^{2} \cdot ....\cdot \left(\rho ^{\kappa } \right)^{2}
\delta ^{\nu } _{\chi } \cdot ....\cdot \delta ^{\kappa } _{\xi }
\end{equation}

\noindent The norm induced from this scalar product  is $L^{m} $
for the corresponding \textit{n}-coordinate tensor.

In summary, plasticity yield criteria can be defined  by the norms
$L^{m} $ induced from the scalar products of stresses
\eqref{GrindEQ__51_} and coordinates \eqref{GrindEQ__52_} in the
stress and coordinate (position) spaces defined by the series
expansion \eqref{GrindEQ__47_}. As a result of this construction,
the space \eqref{GrindEQ__47_} is normed.

\section{Analysis of plasticity yield criteria. Convexity}

The norms defined in Section 9 allow one to assess measurements of
variations of any order with a view to establishing the plasticity
yield criterion for each material in terms of its particular
properties.

\textit{Isotropic criteria} can be analysed by using the norm
$L^{m} $ for  the \textit{first-order variation}
\eqref{GrindEQ__42_}, which gives

\noindent

\begin{equation} \label{GrindEQ__53_}
\begin{array}{l} {\left\| \left. d\sigma \right|_{1} \right\|
^{m} =2\left|\rho ^{3} \right|^{m} \left|\sigma _{yy} -\sigma
_{xx} \right|^{m} +2\left|\rho ^{1} \right|^{m} \left|\sigma _{zz}
-\sigma _{yy} \right|^{m} +2\left|\rho ^{2} \right|^{m}
\left|\sigma _{xx} -\sigma _{zz} \right|^{m} +} \\ {\quad \quad
\quad \quad +\left[2\left|\rho ^{1} \right|^{m} +2\left|\rho ^{2}
\right|^{m} +2^{m+1} \left|\rho ^{3} \right|^{m}
\right]\left|\sigma _{xy} \right|^{m} +\left[2\left|\rho ^{2}
\right|^{m} +2\left|\rho ^{3} \right|^{m} +2^{m+1} \left|\rho ^{1}
\right|^{m} \right]\left|\sigma _{yz} \right|^{m} +} \\ {\quad
\quad \quad \quad +\left[2\left|\rho ^{1} \right|^{m} +2\left|\rho
^{3} \right|^{m} +2^{m+1} \left|\rho ^{2} \right|^{m}
\right]\left|\sigma _{xz} \right|^{m} } \end{array}
\end{equation}

With complete spatial isotropy $(\rho ^{1} =\rho ^{2} =\rho ^{3}
)$  and invariance $(m=2)$, the outcome is the \textit{Huber--Von
Mises criterion} \eqref{GrindEQ__13_}. In the particular case of
using independent plane stresses $xy,yz,xz$ in each coordinate
directions $(0,0,\rho ^{3} )$, $(\rho ^{1} ,0,0)$, $(0,\rho ^{2}
,0)$ respectively, taking $(m=2)$ and factoring in such a way as
to make the criterion smooth (\textit{i.e.}, differentiable) leads
to the \textit{Tresca criterion} \eqref{GrindEQ__12_}. With
complete spatial isotropy $(\rho ^{1} =\rho ^{2} =\rho ^{3} )$,
using the Cauchy tensor in principal stress (Cauchy diagonal
tensor) gives the \textit{Hosford criterion} \eqref{GrindEQ__17_}.
These are the most widely used basic plasticity yield criteria for
materials with completely isotropic symmetry.

\textit{Anisotropic criteria}, which are applicable to
orthotropic materials, can be analysed by using the norm $L^{m} $
for the \textit{second-order variation} \eqref{GrindEQ__44_},
which gives

\begin{equation} \label{GrindEQ__54_}
\begin{array}{l} {\left\| \left. d\sigma \right|_{2} \right\| ^{m}
=2\left|\rho ^{12} \right|^{m} \left|2\sigma _{zz} -\sigma _{xx}
-\sigma _{yy} \right|^{m} +2\left|\rho ^{13} \right|^{m}
\left|2\sigma _{yy} -\sigma _{xx} -\sigma _{zz} \right|^{m}
+2\left|\rho ^{23} \right|^{m} \left|2\sigma _{xx} -\sigma _{yy}
-\sigma _{zz} \right|^{m} +} \\ {\quad \quad \quad +8\left|\rho
^{33} \right|^{m} \left|\sigma _{yy} -\sigma _{xx} \right|^{m}
+8\left|\rho ^{11} \right|^{m} \left|\sigma _{zz} -\sigma _{yy}
\right|^{m} +8\left|\rho ^{22} \right|^{m} \left|\sigma _{xx}
-\sigma _{zz} \right|^{m} +} \\ {\quad \quad \quad +\left[(2+2^{m}
)\cdot 2^{m} \left|\rho ^{12} \right|^{m} +2\cdot 3^{m} \left|\rho
^{23} \right|^{m} +2\cdot 3^{m} \left|\rho ^{13} \right|^{m}
+2\left|\rho ^{11} \right|^{m} +2\left|\rho ^{22} \right|^{m}
+2^{2m+1} \left|\rho ^{33} \right|^{m} \right]\left|\sigma _{xy}
\right|^{m} +} \\ {\quad \quad \quad +\left[2\cdot 3^{m}
\left|\rho ^{12} \right|^{m} +(2+2^{m} )\cdot 2^{m} \left|\rho
^{23} \right|^{m} +2\cdot 3^{m} \left|\rho ^{13} \right|^{m}
+2^{2m+1} \left|\rho ^{11} \right|^{m} +2\left|\rho ^{22}
\right|^{m} +2\left|\rho ^{33} \right|^{m} \right]\left|\sigma
_{yz} \right|^{m} +} \\ {\quad \quad \quad +\left[2\cdot 3^{m}
\left|\rho ^{12} \right|^{m} +2\cdot 3^{m} \left|\rho ^{23}
\right|^{m} +(2+2^{m} )\cdot 2^{m} \left|\rho ^{13} \right|^{m}
+2\left|\rho ^{11} \right|^{m} +2^{2m+1} \left|\rho ^{22}
\right|^{m} +2\left|\rho ^{33} \right|^{m} \right]\left|\sigma
_{xz} \right|^{m} } \end{array}
\end{equation}
\textbf{}

For the particular case of the Cauchy diagonal tensor,  this leads
to \textit{Hill's second criterion} \eqref{GrindEQ__19_}.
Diagonalizing the coordinate tensor $\rho ^{\nu \mu } $ and
assuming $(m=2)$, we are led to \textit{Hill's first criterion}
\eqref{GrindEQ__14_}. On the other hand, if both the coordinate
tensor, $\rho ^{\nu \mu } $, and the Cauchy tensor are
diagonalized, then one obtains the \textit{Logan--Hosford
criterion} \eqref{GrindEQ__18_}. These criteria hold for
anisotropic materials with orthotropic symmetry (\textit{i.e.},
orthotropic materials), a general description of which requires
six spatial parameters (the six parameters of the symmetry tensor
$\rho ^{\nu \mu } $).

The \textit{criteria of Barlat et al.} [2] and \textit{ Karafillis
\& Boyce} [22] were originally established by transforming the
stress tensor; this involved considering symmetry and anisotropy
in the material. Within the framework of the proposed theory, the
transformations required to describe these effects in materials
can be analysed via transformations of the coordinate tensors
$\rho ^{\nu } ,\ldots ,\rho ^{\nu \ldots \kappa } $ from their
starting bases ($X_{\nu } ,\ldots ,X_{\nu ....\kappa } $), to
those needed for an accurate description of the material
concerned. Obviously, anisotropy makes solid materials
basis-dependent (\textit{i.e.}, reference system-dependent);
therefore, it grows with increasing order of the bases used. In
completely isotropic materials, the plasticity yield criterion
depends on the coordinates of a single scalar $(\rho ^{1} =\rho
^{2} =\rho ^{3} )$, \textit{i.e.}, it is absolutely independent of
the particular coordinate system.

In should be noted that \textit{Hill's second criterion},  which
provides an accurate description of the anomalous behaviour of
some materials such as aluminium [18, 37], arises naturally within
the framework of the proposed plasticity theory.\textit{
}Additionally, the proposed theory includes the anisotropy
criterion of Hosford [25] that can explain the behaviour of
\textit{fcc} (for $m=8$) and \textit{bcc} (for $m=6$) materials.

With the norms defined above, equation \eqref{GrindEQ__47_}
describes the most salient plasticity yield criteria reported so
far as particular cases. Therefore, \eqref{GrindEQ__47_} provides
a self-contained description of hydrostatic-pressure independent
plasticity (\textit{i.e.}, plasticity compliant with the proposed
postulates and approximations). A higher-order analysis allows one
to establish a generic criterion in terms of the following
expanded series:

\begin{equation} \label{GrindEQ__55_}
\begin{array}{l} {\left\| d\sigma \right\| ^{m} =\sum _{n{'} =0}
^{\infty }\sum _{f} u_{n{'} f} \left|2^{n{'} } \sigma _{ii}
-(2^{n{'} } -1)\sigma _{jj} -\sigma _{kk} \right|^{m}
+v\left|\sigma _{xy} \right|^{m} +w\left|\sigma _{yz} \right|^{m}
+k\left|\sigma _{xz} \right|^{m} =} \\ {\quad \quad \quad =\ldots
\ldots +u_{a1} \left|2^{a} \sigma _{xx} -(2^{a} -1)\sigma _{yy}
-\sigma _{zz} \right|^{m} +u_{a2} \left|2^{a} \sigma _{xx} -(2^{a}
-1)\sigma _{zz} -\sigma _{yy} \right|^{m} +} \\ {\quad \quad \quad
+u_{a3} \left|2^{a} \sigma _{yy} -(2^{a} -1)\sigma _{xx} -\sigma
_{zz} \right|^{m} +u_{a4} \left|2^{a} \sigma _{yy} -(2^{a}
-1)\sigma _{zz} -\sigma _{xx} \right|^{m} +} \\ {\quad \quad \quad
+u_{a5} \left|2^{a} \sigma _{zz} -(2^{a} -1)\sigma _{xx} -\sigma
_{yy} \right|^{m} +u_{a6} \left|2^{a} \sigma _{zz} -(2^{a}
-1)\sigma _{yy} -\sigma _{xx} \right|^{m} +} \\ {\quad \quad \quad
+\ldots \ldots \ldots \ldots \ldots \ldots \ldots \ldots +u_{21}
\left|4\sigma _{xx} -3\sigma _{zz} -\sigma _{yy} \right|^{m}
+u_{22} \left|4\sigma _{xx} -3\sigma _{yy} -\sigma _{zz}
\right|^{m} +} \\ {\quad \quad \quad +u_{23} \left|4\sigma _{yy}
-3\sigma _{zz} -\sigma _{xx} \right|^{m} +u_{24} \left|4\sigma
_{yy} -3\sigma _{xx} -\sigma _{zz} \right|^{m} +u_{25}
\left|4\sigma _{zz} -3\sigma _{xx} -\sigma _{yy} \right|^{m} +} \\
{\quad \quad \quad +u_{26} \left|4\sigma _{zz} -3\sigma _{yy}
-\sigma _{xx} \right|^{m} +u_{11} \left|2\sigma _{zz} -\sigma
_{xx} -\sigma _{yy} \right|^{m} +u_{12} \left|2\sigma _{yy}
-\sigma _{xx} -\sigma _{zz} \right|^{m} +} \\ {\quad \quad \quad
+u_{13} \left|2\sigma _{xx} -\sigma _{yy} -\sigma _{zz}
\right|^{m} +u_{01} \left|\sigma _{yy} -\sigma _{xx} \right|^{m}
+u_{02} \left|\sigma _{zz} -\sigma _{yy} \right|^{m} +u_{03}
\left|\sigma _{xx} -\sigma _{zz} \right|^{m} +} \\ {\quad \quad
\quad +v\left|\sigma _{xy} \right|^{m} +w\left|\sigma _{yz}
\right|^{m} +k\left|\sigma _{xz} \right|^{m} } \end{array}
\end{equation}

\noindent where $i\ne j\ne k\ne i$ and the indices can range  over
$x,y,z$; $f$ is the permutation index. Each term in
\eqref{GrindEQ__55_} has an associated coefficient $u_{n{'} f}
,v,w,k$ that can be explicitly computed within the framework of
the proposed unified theory, but has been excluded in the expanded
from \eqref{GrindEQ__55_} for simplicity. Note that positive
integer parameter $n=n{'} +1$ in eq. \eqref{GrindEQ__55_}
represents the \textit{n}-order approximation.

Stress convexity in \eqref{GrindEQ__55_} can be assessed by
assuming that a linear combination of convex functions with
positive coefficients is a convex function itself [36]. The
coefficients $u_{n{'} f} ,v,w,k$ are in fact all positive.
Therefore, it will suffice to show that the stress-dependence of
each individual coefficient in \eqref{GrindEQ__55_} is convex
(\textit{i.e.}, that all stresses in the terms $\left|2^{n{'} }
\sigma _{ii} -(2^{n{'} } -1)\sigma _{jj} -\sigma _{kk} \right|^{m}
$, $\left|\sigma _{xy} \right|^{m} $, $\left|\sigma _{yz}
\right|^{m} $, $\left|\sigma _{xz} \right|^{m} $ are convex).
Convexity in a function \textit{T} of three variables can be
expressed [36] as

\begin{equation} \label{GrindEQ__56_}
T(c_{1} \sigma _{ii} +c_{2} \sigma _{jj} +c_{3} \sigma _{kk} )\le
c_{1} T(\sigma _{ii} )+c_{2} T(\sigma _{jj} )+c_{3} T(\sigma _{kk}
)
\end{equation}

\noindent where $c_{1} ,c_{2} ,c_{3} \in \Re $ are  positive or
zero and the following condition holds: $c_{1} +c_{2} +c_{3} =1$.
The convexity of $\left|\sigma _{xy} \right|^{m} $, $\left|\sigma
_{yz} \right|^{m} $, $\left|\sigma _{xz} \right|^{m} $ can be
immediately confirmed by applying \eqref{GrindEQ__56_} to the
particular case of a single variable. That of $\left|2^{n{'} }
\sigma _{ii} -(2^{n{'} } -1)\sigma _{jj} -\sigma _{kk} \right|^{m}
$ can also be easily inferred from the triangular inequality (see
Appendix) fulfilling this $L^{m} $ norm as defined in the stress
vector space $(\sigma _{xx} ,\sigma _{yy} ,\sigma _{zz} )$. Note
that, for $L^{m} $ to actually be the norm, $m\in \Re ,^{} m\ge
1$, which is the constraint to be imposed on \eqref{GrindEQ__55_}
for stresses to be convex. Under these conditions, the plasticity
yield criterion \eqref{GrindEQ__55_} will fulfil the \textit{third
postulate}. Within the framework of the proposed theory, the
general criterion \eqref{GrindEQ__55_} can predict plastic
behaviour in solids. Therefore, it can be of use in future
experimental and theory studies on this topic.

\noindent

\section{Hydrostatic pressure-dependent plasticity yield criteria}

The hydrostatic  pressure-independent definitions obtained in the
previous sections invariably include traceless 2-tensors. In fact,
this mathematical constraint ensures that the ensuing criteria
will be hydrostatic-pressure independent. As noted earlier, the
normal space to the Cauchy tensor describes the hydrostatic
pressure-dependence of plasticity. As a result, considering the
effects of hydrostatic pressure in plasticity yield criteria
entails including a $c\left(\sigma _{M} \right)^{b} \equiv
c\left(\sigma _{xx} +\sigma _{yy} +\sigma _{zz} \right)^{b} $
dependent function to describe the hydrostatic behaviour of the
material with specific coefficients $b,c$ [10, 11, 7, 9]. A more
detailed study of the hydrostatic pressure-dependence in the
framework of unified theory will be the subject of future work.

\noindent

\section{Conclusions}

In this paper we propose a new approach to the theory of
macroscopic plasticity, based on a geometrical ansatz, that
unifies the different plasticity yield criteria. The method is
based on decomposing the Cauchy tensor into its component stress
spaces: tangent and normal spaces. Hydrostatic-pressure
independent yield criteria are located on the tangent space to the
Cauchy tensor (TSCT), whereas hydrostatic-pressure dependent yield
criteria are located on the tangent and normal spaces to the
Cauchy tensor (TSCT and NSCT). The analysis of the corresponding
spaces requires the use of Lie groups of transformations.

This work focuses on the tangent space to Cauchy tensor (TSCT) and
analyzes the hydrostatic-pressure independent yield criteria as
orthogonal transformations. The measure of this tangent space is
taken as a mathematical norm that guarantees the convexity of the
yield criteria. From this measure we obtained the convex-stress
dependence for both isotropic yield criteria (Tresca 1864 and Von
Mises 1913) and anisotropic yield criteria (Hill 1948 and Hill
1979). In addition, a stress series expansion is obtained that
allows a detailed study of the mechanical properties in materials:
anisotropy, hardening-softening, etc. The tensor coefficients
resulting from orthogonal transformations are dependent on
internal variables. These tensor coefficients allow a deep
analysis of mechanical properties in materials. This unified
theory simplifies the great existing amount of yield criteria and
gives a physical meaning to the anisotropy coefficients.

This unified  theory also allows an analytical study of anisotropy
(dependency of $f$on $\xi $) and hardening-softening (dependency
of $\dot{f}$ on $\xi $) by means of a set of tensor coefficients.
These coefficients are the directions of macroscopic plasticity
(by points) in materials. A connection is suggested between
macroscopic transformations (unified theory) and the microscopic
processes (dislocation theory). To this extent, the concept of
macroscopic point, that constitutes a statistical measure of
microscopic processes, is introduced. The concept of the point
connects two scales with different physical laws.

This unified  approach has a series of advantages. At first, it
proposes the stress dependencies of yield criteria, starting from
very simple postulates and with a unified vision of the underlying
phenomena. Further, it allows to know the macroscopic plasticity
directions in the solid points, by tensor coefficients of
orthogonal transformations. It also provides a manner to analyze
the anisotropy and hardening-softening in materials. Finally, it
enables us to establish a connection between microscopic
plasticity and macroscopic plasticity.

In future work it is important to extend the unified theory to
treat the very important question about the hydrostatic
pressure-dependent plasticity (important in composites and
granular materials); for this it is necessary to analyse
dilatation transformations. In this context, it is of fundamental
importance to make a deep study of hardening-softening and the
time-evolution of plasticity in materials; for the latter aspect
it is necessary to determine an adequate time-transformation that
gives a correct parameterization of the solid time-evolution.

\newpage

\section*{Notation}

\subsection*{(a) Indices}

\begin{enumerate}
\item Solid point position: $\lambda $ ranging over $x,y,z$

\item  Cauchy tensor: $i,j,k,l,p,q$ ranging over $x,y,z$

\item  Rotation group: $\nu ,\mu ,\omega ,\kappa ,\chi ,\xi $ ranging over $1,2,3$

\item  Levi--Civita tensor: $a,b,c$ ranging over $1,2,3$

\item  Elastic--plastic limit: \textit{P}

\item  Angle equivalence: $(\theta ^{1} ,\theta ^{2} ,\theta ^{3}
)\equiv (\alpha ,\beta ,\gamma )$

\item  Cauchy tensor equivalence:  $\sigma \equiv \sigma ^{kl} $
(contravariant); $\sigma \equiv \sigma _{kl} $ (covariant)

\item  Principal stresses of the Cauchy tensor: $\sigma _{1} ,\sigma _{2} ,\sigma _{3} $

\item  Hydrostatic pressure: $\sigma _{M} $

\item  Internal variables: $\xi =\xi _{\alpha } $
\end{enumerate}

\subsection*{(b) Tensors}

\noindent The Einstein summation  convention is used throughout.
Covariant tensor notation is systematically employed from Section
4 onwards.

\subsection*{Acknowledgements. } The first author, Jos\'e Miguel Luque
Raig\'on, is grateful to Professor Alfredo Navarro of the
University of Seville for his helpful, inspiring comments on
materials mechanics and plasticity, and also to the members of the
Mechanical Engineering and Materials Research Group of the
University of Seville for their support. This work was funded by
the Andalusian regional government (Junta de Andaluc\'{\i}a)
within the framework of Project P06-TEP-01752.\textit{}

\newpage

\section*{Appendix: Convexity of general plasticity function}

\noindent In order to prove the convexity of the plasticity
function (section 10), it is necessary to show the convexity of
the terms $\left|2^{n{'} } \sigma _{ii} -(2^{n{'} } -1)\sigma
_{jj} -\sigma _{kk} \right|^{m} $ with $m\ge 1$, $m\in \Re $ and
\textit{$n{'} $} a\textit{ }positive\textit{ } integer or zero. If
the triangular inequality is taken we obtain

$\left|2^{n{'} } \sigma _{ii} -\left(2^{n{'} } -1\right) \sigma
_{jj} -\sigma _{kk} \right|^{m} \le 2^{n{'} \cdot m} \left|\sigma
_{ii} \right|^{m} +\left(2^{n{'} } -1\right)^{m} \left|\sigma
_{jj} \right|^{m} +\left|\sigma _{kk} \right|^{m} $ (A)

\noindent

\noindent Dividing (A) by $2^{m(n{'} +1)} $ results in

\noindent

$\begin{array}{c} {\left|\frac{1}{2} \sigma _{ii}
-\left(\frac{1}{2} -\frac{1}{2^{n{'} +1} } \right)\sigma _{jj}
-\frac{1}{2^{n{'} +1} } \sigma _{kk} \right|^{m} \le
\frac{1}{2^{m} } \left|\sigma _{ii} \right|^{m} +\left(\frac{1}{2}
-\frac{1}{2^{n{'} +1} } \right)^{m} \left|\sigma _{jj} \right|^{m}
+} \\ {\frac{1}{2^{m(n{'} +1)} } \left|\sigma _{kk} \right|^{m}
\le \frac{1}{2} \left|\sigma _{ii} \right|^{m} +\left(\frac{1}{2}
-\frac{1}{2^{n{'} +1} } \right)\left|\sigma _{jj} \right|^{m}
+\frac{1}{2^{n{'} +1} } \left|\sigma _{kk} \right|^{m} }
\end{array}$ (B)

\noindent

Thus taking $T(x)\equiv \left|x\right|^{m} $, the condition of the
convexity is fulfilled

$T(c_{1} \sigma _{ii} +c_{2} \sigma _{jj} +c_{3} \sigma _{kk} )\le
c_{1} T(\sigma _{ii} )+c_{2} T(\sigma _{jj} )+c_{3} T(\sigma _{kk}
)$ (C),

where $c_{1} =\frac{1}{2} $;  $c_{2} =\frac{1}{2}
-\frac{1}{2^{n{'} +1} } $; $c_{3} =\frac{1}{2^{n{'} +1} } $ and
\textit{$n{'} $} is a positive integer or zero; with $c_{1} ,c_{2}
,c_{3} \ge 0$ and $c_{1} +c_{2} +c_{3} =1$. So the convexity of
the plasticity function has been proved.

\newpage

\section*{Figure captions}

\noindent \textbf{Figure 1.} Reference systems for the overall
solid $\{ O,x^{\lambda } =\left({\rm x,\; y,\; z}\right)\} $ and
for each point \textit{p} in it $\{ O_{p} ,x^{\lambda }
=\left({\rm x,\; y,\; z}\right)\} $. Each point has an associated
Cauchy tensor acting on it (see Figure~2).

\noindent \textbf{}

\noindent \textbf{Figure 2.}  Representation of the Cauchy tensor
$\sigma ^{kl} $ and its rotational possibilities about the Euler
angles $\Theta _{E} =\Theta _{E} (\alpha ,\beta ,\gamma )$ in the
reference system ${\rm \{ O}_{{\rm p}} {\rm ,x}^{\lambda } \} $.
The sphere surface in the figure is a particular making of the
$SO\eqref{GrindEQ__3_}$ rotation group.

\noindent \textbf{}

\noindent \textbf{Figure 3.}  Action of the
$SO\eqref{GrindEQ__3_}$ rotation group on the Cauchy tensor
$\sigma ^{kl} $ for a partial rotation $\beta $ about the
\textit{$y$-}axis, represented by the reference system
transformation ${\rm \{ O}_{{\rm p}} {\rm ,x}^{\lambda } \} \to
{\rm \{ O}_{{\rm p}} {\rm ,}\bar{{\rm x}}^{\lambda } \} $.

\noindent \textbf{}

\noindent \textbf{Figure 4.}  Arbitrary representation of TSCT as
an ellipsoid. The tangent space $TSO\eqref{GrindEQ__3_}_{\rho
^{\mu } } $ that provides a first-order approximation of TSCT to a
reduced neighbourhood of  $\rho ^{\mu } $ is shown.

\noindent \textbf{}

\noindent \textbf{Figure 5.} Arbitrary representation of TSCT and
its approximations of different order to an infinitesimal
neighbourhood. The higher orders lead to local infinitesimal
neighbourhood that provide a more accurate description of the
overall structure of TSCT.

\newpage

\begin{figure}
\includegraphics{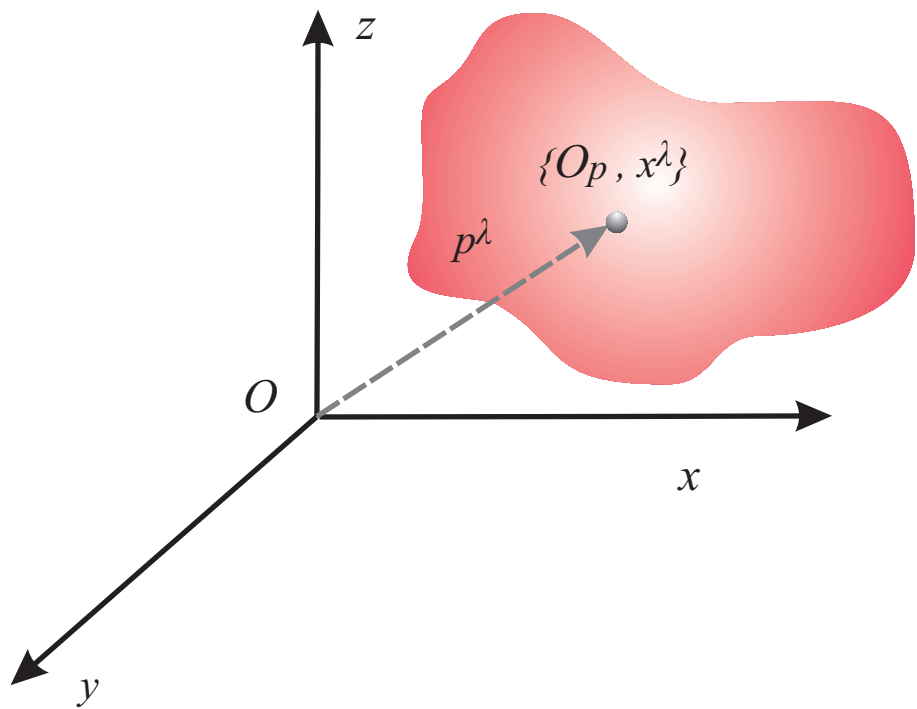}\\
\caption{Reference systems for the overall solid $\left\{
O,x^{\lambda}\right\} =\left(\left\{{\rm x,\; y,\;
z}\right\}\right)\}$ and for each point $p$ in it $\left\{
O_{p},x^{\lambda}\right\} =\left(\left\{{\rm x,\; y,\;
z}\right\}\right)\}$. Each point has an associated Cauchy tensor
acting on it (see Figure 2).}\label{Fig1}
\end{figure}

\noindent

\begin{figure}
\includegraphics{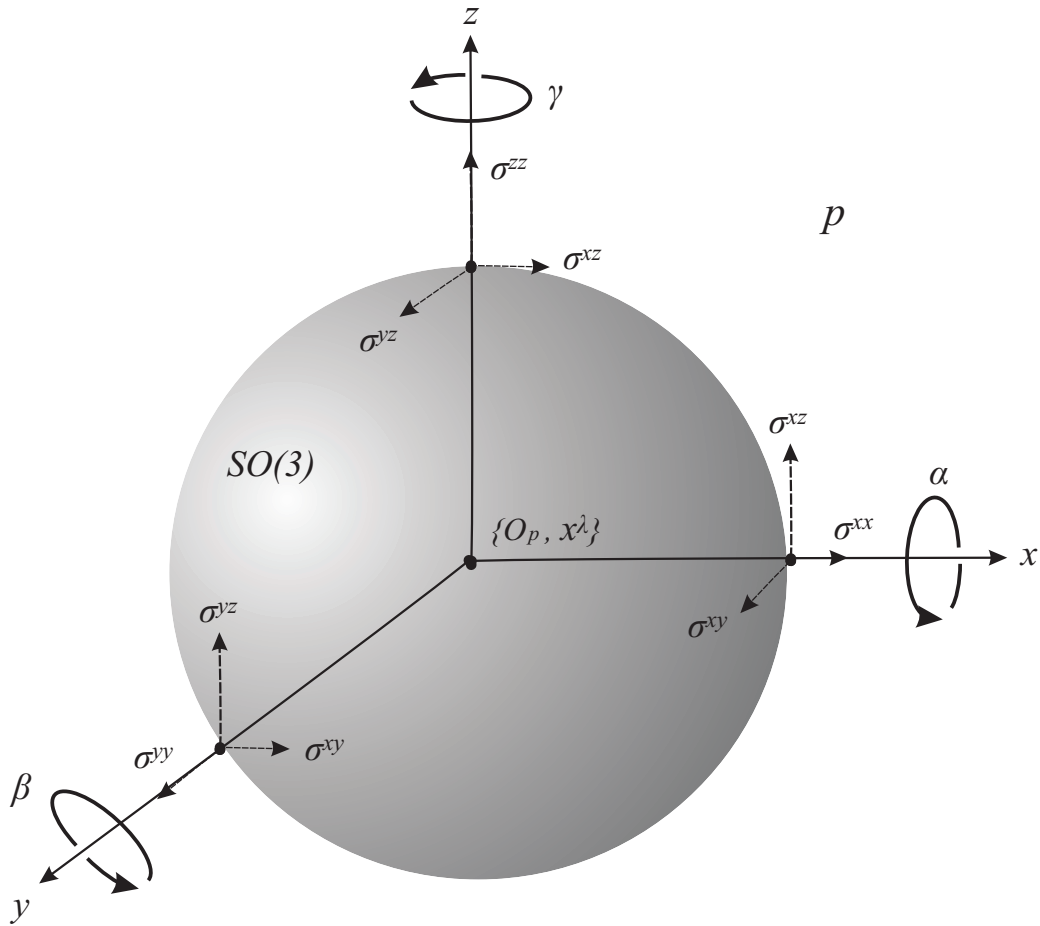}\\
\caption{Representation of the Cauchy tensor $\sigma ^{kl} $ and
its rotational possibilities about the Euler angles $\Theta _{E}
=\Theta _{E} (\alpha ,\beta ,\gamma )$ in the reference system
${\rm \{ O}_{{\rm p}} {\rm ,x}^{\lambda } \} $. The sphere surface
in the figure is a particular making of the $SO\eqref{GrindEQ_3_}$
rotation group.}\label{Fig2}
\end{figure}

\begin{figure}
\includegraphics{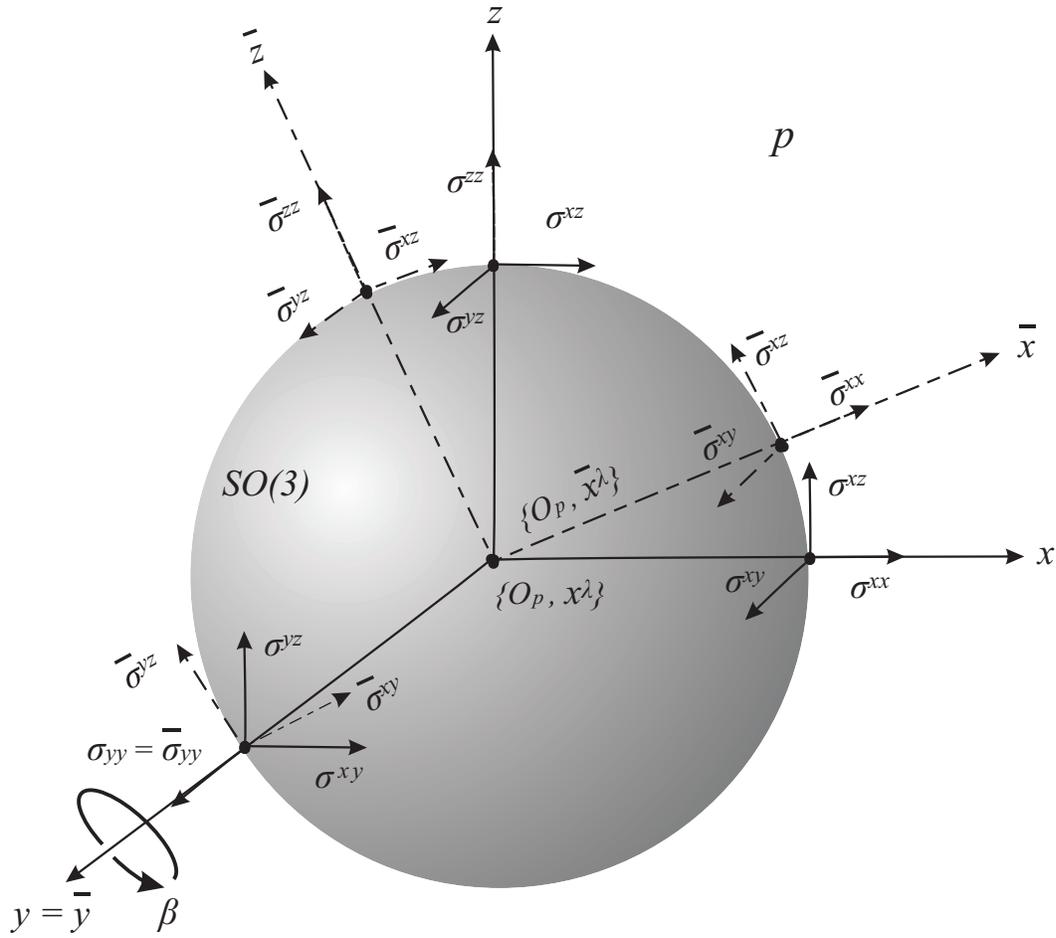}\\
\caption{Action of the $SO\eqref{GrindEQ__3_}$ rotation group on
the Cauchy tensor $\sigma ^{kl} $ for a partial rotation $\beta $
about the \textit{$y$}-axis, represented by the reference system
transformation ${\rm \{ O}_{{\rm p}} {\rm ,x}^{\lambda } \} \to
{\rm \{ O}_{{\rm p}} {\rm ,}\bar{{\rm x}}^{\lambda } \}
$.}\label{Fig3}
\end{figure}

\noindent

\begin{figure}
\includegraphics{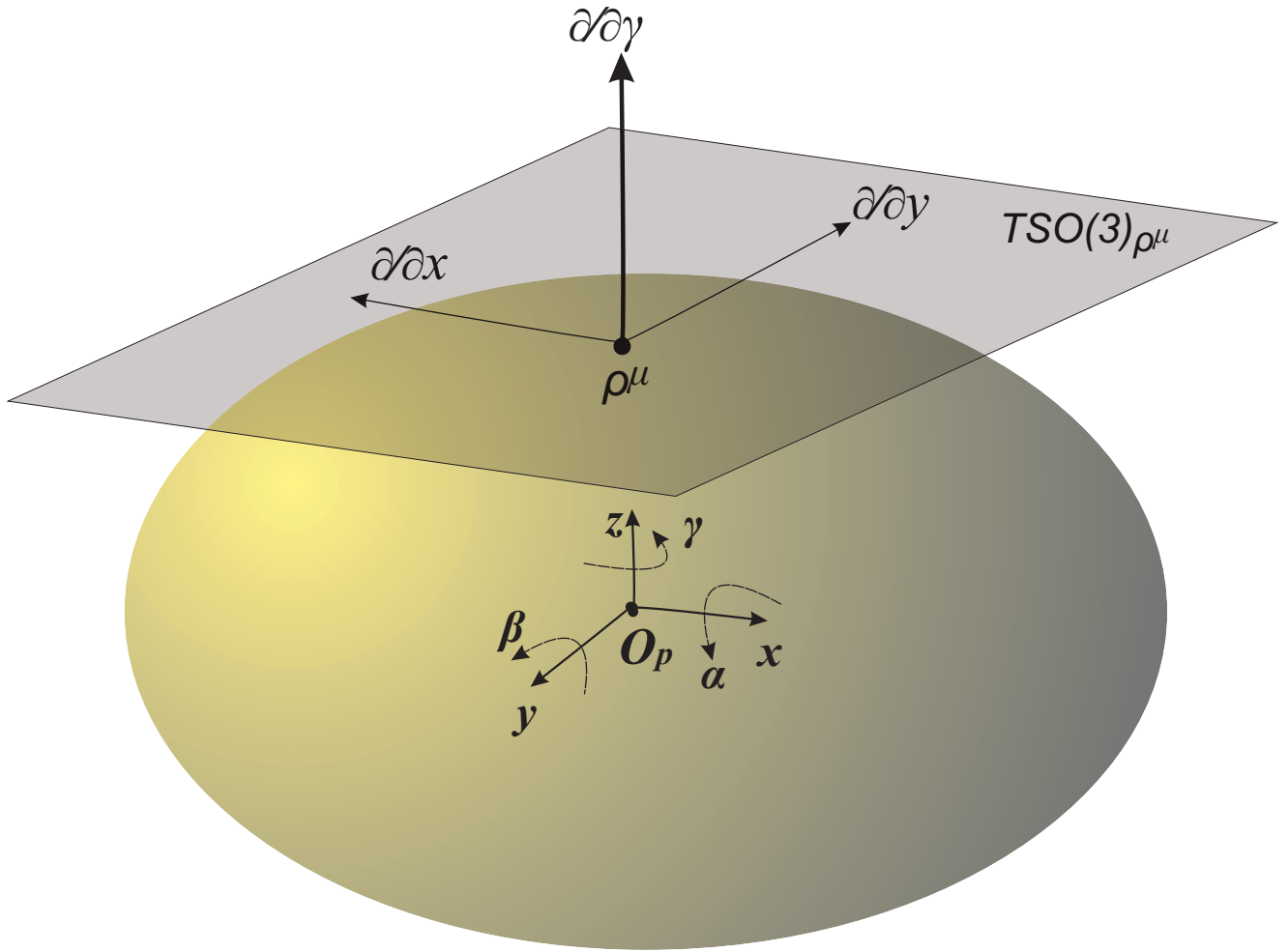}\\
\caption{Arbitrary representation of TSCT as an ellipsoid. The
tangent space $TSO\eqref{GrindEQ__3_}_{\rho ^{\mu } } $ that
provides a first-order approximation of TSCT to a reduced
neighbourhood of $\rho ^{\mu } $ is shown.}\label{Fig4}
\end{figure}

\noindent

\begin{figure}
\includegraphics{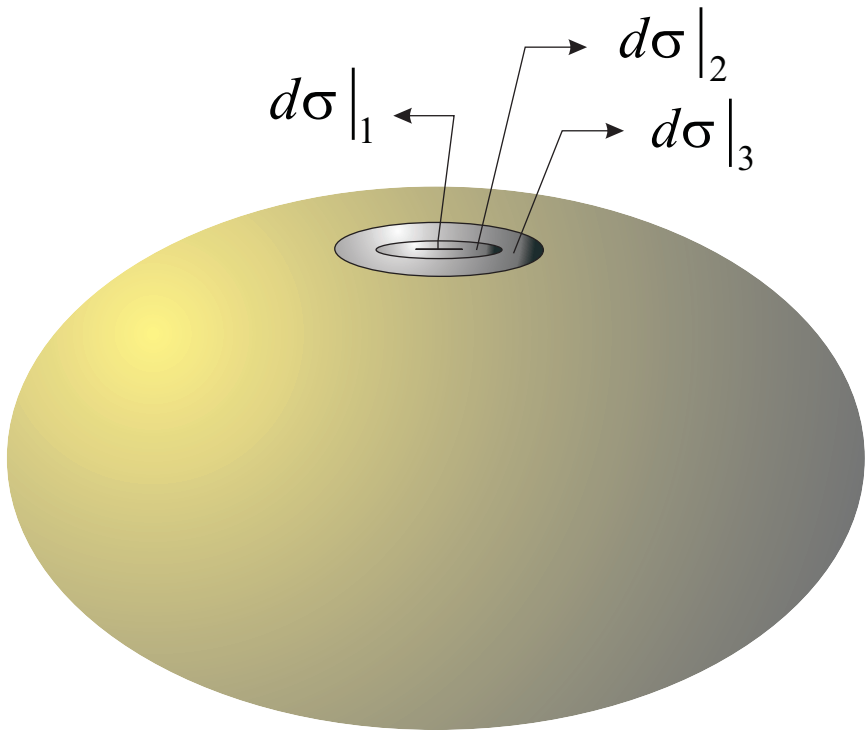}\\
\caption{Arbitrary representation of TSCT and its approximations
of different order to an infinitesimal neighbourhood. The higher
orders lead to local infinitesimal neighbourhood that provide a
more accurate description of the overall structure of
TSCT.}\label{Fig5}
\end{figure}

\end{document}